\documentclass[aps,prb,twocolumn,groupedaddress,floatfix,longbibliography,superscriptaddress]{revtex4-2}

\usepackage{amsmath}
\usepackage{amssymb}
\usepackage{amsfonts}
\usepackage{bbold}
\usepackage{bm}
\usepackage{cancel}
\usepackage{times,float}
\usepackage{graphicx}
\usepackage[usenames,dvipsnames,svgnames]{xcolor}
\usepackage{hyperref}
\hypersetup{colorlinks=true, linkcolor=Blue, citecolor=Blue,urlcolor=Blue}
\usepackage{multirow}
\usepackage{ulem}
\usepackage{color,epsfig}
\usepackage{slashed}
\usepackage{feynmf}
\usepackage{array}
\usepackage{physics}
\usepackage{subcaption}
\usepackage{soul}
\usepackage{bm}
\usepackage{graphicx}
\graphicspath{{FIGURES/}}
\usepackage{bbold}
\usepackage{cancel}
\usepackage{mathrsfs}
\usepackage[compatibility=false]{caption}
\usepackage{times,float}
\usepackage{physics}
\usepackage{caption}
\usepackage{xcolor}
\usepackage{subcaption}

\newcommand\ra{\rightarrow}

\begin{document}

\title{Thermoelectric transport in graphene under strain fields modeled by Dirac oscillators}

\author{Juan A. Ca\~nas}
\email{juan.canas@correo.nucleares.unam.mx}
\address{Instituto de Ciencias Nucleares, Universidad Nacional Aut\'{o}noma de M\'{e}xico, 04510 Ciudad de M\'{e}xico, M\'{e}xico}

\author{Daniel A. Bonilla}
\email{daniel.bonillam@correo.nucleares.unam.mx}
\address{Instituto de Ciencias Nucleares, Universidad Nacional Aut\'{o}noma de M\'{e}xico, 04510 Ciudad de M\'{e}xico, M\'{e}xico}

\author{A. Mart\'{i}n-Ruiz}
\email{alberto.martin@nucleares.unam.mx}
\address{Instituto de Ciencias Nucleares, Universidad Nacional Aut\'{o}noma de M\'{e}xico, 04510 Ciudad de M\'{e}xico, M\'{e}xico}

\begin{abstract}
Graphene has emerged as a paradigmatic material in condensed matter physics due to its exceptional electronic, mechanical, and thermal properties. A deep understanding of its thermoelectric transport behavior is crucial for the development of novel nanoelectronic and energy-harvesting devices. In this work, we investigate the thermoelectric transport properties of monolayer graphene subjected to randomly distributed localized strain fields, which locally induce impurity-like perturbations. These strain-induced impurities are modeled via 2D Dirac oscillators, capturing the coupling between pseudorelativistic charge carriers and localized distortions in the lattice. Employing the semiclassical Boltzmann transport formalism, we compute the relaxation time using a scattering approach tailored to the Dirac oscillator potential. From this framework, we derive analytical expressions for the electrical conductivity, Seebeck coefficient, and thermal conductivity. The temperature dependence of the scattering centers density is also investigated. Our results reveal how strain modulates transport coefficients, highlighting the interplay between mechanical deformations and thermoelectric performance in graphene. This study provides a theoretical foundation for strain engineering in thermoelectric graphene-based devices.
\end{abstract}

\maketitle

%-----------------------------------------------------------

\section{Introduction}

\noindent

Since its experimental isolation in 2004 \cite{Novoselov_2004}, graphene has captivated the condensed matter community due to its extraordinary physical properties, such as high carrier mobility \cite{Novoselov_2004,BOLOTIN2008351}, mechanical strength \cite{Sc_Lee_2008,PSS_Lee_2009}, and thermal conductivity \cite{NL_Balandin_2008,NA_Balandin_2011}. As the first truly two-dimensional (2D) material, graphene has also paved the way for the exploration of a wide range of other 2D systems, including transition metal dichalcogenides \cite{NRM_Manzeli_2017}, hexagonal boron nitride \cite{AM_Roy_2021}, and layered topological materials \cite{AFM_Yu_2025}. Among these, graphene remains a prime candidate for next-generation technologies, particularly in nanoelectronics, flexible devices, and thermoelectric energy conversion \cite{G_Radsar_2021,C_Miao_2023,CC_Kim_2025}. Understanding the thermoelectric transport properties of graphene is thus essential not only for optimizing its performance in practical applications but also for gaining fundamental insight into carrier dynamics in low-dimensional systems. The ability to manipulate and enhance these properties through external means such as strain or disorder offers a promising route toward functional device design. For instance, thermoelectric transport in Dirac materials such as graphene and Weyl semimetals has recently been studied in models with quenched disorder arising from a random distribution of point-like or structural defects in the material. For example, this has been analyzed in monolayer graphene coupled to a three-dimensional topological insulator~\cite{Bonilla_Martin}, and in Weyl semimetals with screw dislocation-type defects, where strain-induced pseudomagnetic fields also emerge~\cite{Soto-Garrido_2018,Soto_2,Bonilla_1,Bonilla_2,Bonilla_3}. These studies suggest that the engineering of such materials with point or structural defects represents a promising avenue for the design of thermoelectric materials with enhanced performance.

Recent studies have shown that strain engineering in graphene can be employed to tailor its electronic spectrum, mimicking gauge fields that act on charge carriers as if they were subjected to magnetic fields \cite{E_Guinea_2010,G_Guinea_2010,S_Levy_2010,P_Ramezani_2013,G_de_2013, N_Si_2016,E_G_2017,Munoz_2017,AM_McRae_2024}. In a more complex scenario, however, strain fields are nonuniform and randomly distributed, giving rise to localized perturbations that can significantly affect transport properties \cite{VOZMEDIANO2010109}. These strain-induced distortions can be interpreted as effective impurities, modifying the motion of Dirac fermions and leading to scattering processes that impact both charge and heat conduction. In this context, strain-induced gauge fields provides a powerful theoretical framework to model such localized strain effects, as it captures the coupling between pseudorelativistic particles and a confining potential that resembles the impact of local deformations. Beyond their effect on transport, strain fields can also influence the valley degree of freedom \cite{PRL_Settnes_2016,Munoz_2017, PRB_Ortiz_2022}, a key quantity in valleytronics, where electronic information is encoded in the momentum-space valleys of the band structure \cite{CPB_Zheng__2024}. Understanding how such strain-induced perturbations affect valley-dependent transport is therefore essential for advancing valley-based logic and information processing devices.

In this work, we present a theoretical study of thermoelectric transport in strained graphene, where randomly distributed local deformations are modeled using the Dirac oscillator framework \cite{INC_It0_1967,T_Moshinsky_1989}. We begin by demonstrating the emergence of pseudogauge fields induced by strain in the low-energy effective theory of graphene, which act analogously to magnetic fields on Dirac fermions. These pseudomagnetic fields provide a natural setting to study the impact of localized strain through a confining potential. Using semiclassical kinetic theory, we derive expressions for the thermoelectric transport coefficients \cite{ziman_principles}, namely, the electrical conductivity, Seebeck coefficient, and thermal conductivity, under the influence of such randomly distributed strain fields. The relaxation time entering the Boltzmann formalism is computed via a scattering approach, where the Dirac oscillator potential serves as an effective impurity model. We obtain analytical and numerical results for the transport coefficients as functions of the Fermi energy and the pseudomagnetic field strength. To capture finite-temperature behavior, we apply the Sommerfeld expansion, allowing us to explore the thermal response of the system in experimentally relevant regimes. Our findings shed light on the interplay between strain and thermoelectric performance, offering potential strategies for controlling energy and valley transport in graphene-based devices.

The outline of this work is the following. In Section \ref{Model} we introduce in detail the model (strained graphene) and demonstrate that a strain-induced constant pseudomagnetic field can be modeled by a relativistic 2D Dirac oscillator. We also discuss the pseudo-Landau levels. In Section \ref{Transport_Section} we derive general expressions for the transport coefficients within a kinetic theory approach, and using a scattering analysis, we evaluate the transport time. Using realistic parameter values, in Section \ref{Results} we present the main results of this paper. Finally, in Section \ref{Conclusions}, we summarize our results and provide some context for the relevance of our findings.

\section{The model: strain-induced pseudogauge fields in graphene} \label{Model}

\noindent

\begin{figure}
    \centering
    \includegraphics[width=0.5\linewidth]{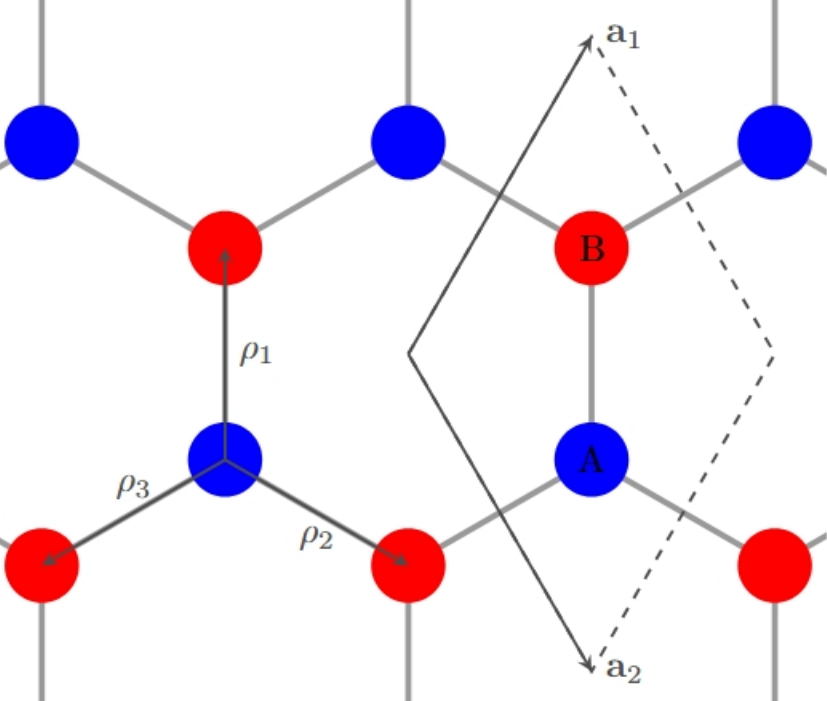}
    \caption{Graphene crystalline structure. Blue and red atoms  the two distinct sub-lattices (A and B, respectively). The Bravais lattice vectors ($\bm{a}_1$, $\bm{a}_2$) and the nearest-neighbor vectors ($\bm{\rho}_1$, $\bm{\rho}_2$, $\bm{\rho}_3$) connecting sub-lattice A to sub-lattice B are also shown.} 
    \label{fig:Graphene_Lattice}
\end{figure}

Graphene is composed of carbon atoms arranged in a two-dimensional hexagonal honeycomb lattice, the side length of the hexagon  is $a\approx 0.142$ nm \cite{RevModPhys.81.109}. Structurally, it can also be described as two interpenetrating triangular sub-lattices (denoted as A and B), as depicted in Fig. \ref{fig:Graphene_Lattice}. The atomic positions in sub-lattice A are given by $\bm{R}_{n} = n_{1} \bm{a}_{1} + n_{2} \bm{a}_{2}$, where \cite{B_Gorbar2021} the primitive lattice vectors are
\begin{equation}
    \bm{a}_{1} = \left( \frac{\sqrt{3} a}{2}, \frac{3 a}{2} \right), \quad \bm{a}_{2} = \left( \frac{\sqrt{3} a}{2}, -\frac{3 a}{2} \right).
\end{equation}
Here $n_{1}$ and $n_{2}$ are integer indices labeling the lattice sites, and the subscript $n$ refers to the $n$-th site in sub-lattice A. The nearest-neighbor positions for an atom in sub-lattice A are described by the vectors $\bm{\rho}_{1}$, $\bm{\rho}_{2}$ and $\bm{\rho}_{3}$ (see Fig.~\ref{fig:Graphene_Lattice}) defined explicitly as 
\begin{equation}
    \bm{\rho}_{1} = \left( 0, a \right), \,\, \bm{\rho}_{2} = \left( \frac{\sqrt{3} a}{2}, - \frac{a}{2} \right), \,\,\bm{\rho}_{3} = \left( -\frac{\sqrt{3} a}{2}, -\frac{a}{2} \right).
    \label{NearNeig_Vecs}
\end{equation}
The tight-binding Hamiltonian for graphene, considering only nearest-neighbor interactions, takes the form
\begin{equation}
    \hat{\mathcal{H}} = - \sum_{\bm{R}_n} \sum_{j=1}^{3} t_{j} \hat{a}^{\dagger}_{\bm{R}_n} \hat{b}_{\bm{R}_n + \bm{\rho}_j } + \mbox{H.c.}
    \label{TB_Ham}
\end{equation}
where $t_j$ denotes the hopping parameter between nearest-neighbor sites, and $\mbox{H.c.}$ stands for the Hermitian conjugate. The operators  $\hat{a}^{\dagger}_{\bm{R}_{n}}$ and $\hat{b}_{\bm{R}_{n} + \bm{\rho}_{j}}$ create and annihilate electrons at sites in sub-lattices A and B, respectively. Diagonalization of Eq. (\ref{TB_Ham}) is achieved by introducing the Fourier transform operators $\hat{a}_{\bm{k}} = (\sqrt{N})^{-1}\sum_{\bm{R}_n} \hat{a}_{\bm{R}_n} e^{-i \bm{k} \cdot \bm{R}_n}$, with $N$ the number of unit cells in the material. Applying the standard algebraic properties of these operators, yields
\begin{equation}
    \hat{\mathcal{H}} = \sum_{\bm{k}\in FBZ} \left( \hat{a}^{\dagger}_{\bm{k}} , \hat{b}^{\dagger}_{\bm{k}} \right) \left( \begin{array}{cc}
       0  & f_{\bm{k}} \\
       f^{*}_{\bm{k}}  & 0
    \end{array} \right) \left( \begin{array}{c}
       \hat{a}_{\bm{k}} \\
       \hat{b}_{\bm{k}}
    \end{array} \right),
    \label{TB-GHam}
\end{equation}
where $f_{\bm{k}} = -\sum_{j=1}^{3} t_{j} e^{i \bm{k} \cdot \bm{\rho}_j}$. This Hamiltonian describes a collection of non-interacting $\bm{k}$-dependent modes, each governed by the $\bm{k}$-dependent Hamiltonian
\begin{equation}
    \hat{\mathcal{H}}(\bm{k}) = \left( \begin{array}{cc}
       0  & f_{\bm{k}} \\
       f^{*}_{\bm{k}}  & 0
    \end{array} \right).
    \label{k-dep_Ham}
\end{equation}
Using the nearest-neighbor vectors defined in Eq.~(\ref{NearNeig_Vecs}), the matrix element $f_{\bm{k}}$ takes the explicit form
 \begin{equation}
     f_{\bm{k}} = -\left( t_1 e^{i a k_y} + t_2 e^{i \frac{a}{2} \left( \sqrt{3} k_x - k_y \right)} + t_3 e^{ -i \frac{a}{2} \left( \sqrt{3} k_x + k_y \right)} \right).
    \label{f_k-general}
 \end{equation}
 
 In pristine graphene it is assumed that all of the hoping parameters are the same ($t_j=t$), with $t\approx 2.97$ eV  obtained from fitting experimental or first-principles data \cite{T_Reich_2002}. Analysis of the dispersion relationship reveals six points where the conduction and valence band touch. However, owing to the lattice symmetries, only two of these points, known as valleys or Dirac points $\bm{K}_D^{\pm} = \left( \pm 4\pi(3\sqrt{3}a)^{-1},0 \right)$, are physically inequivalent. Near the Dirac points, $f_{\bm{k}}$ can be expanded to linear order by writing $\bm{k} = \bm{K}_D^{\pm} + \bm{q}$, where $|\bm{q}| \ll |\bm{K}_D^{\pm}|$ is a small momentum deviation. This approximation yields the matrix element 
 \begin{equation}
     f_{\bm{K}_D^{\pm} + \bm{q}} = \approx  \pm \hbar v_F \left( q_x \mp i q_y  \right),
    \label{f_k-lowE}
 \end{equation}
where $v_F = \frac{3at}{2\hbar} \sim 9.8 \times 10 ^{5}$ m/s is the Fermi velocity. The resulting form transforms Eq.~(\ref{k-dep_Ham}) into a massless Dirac Hamiltonian for each valley. However, to obtain the correct low-energy description expression for the $\bm{K}_D^{-}$ valley, one must reverse the order of creation/annihilation operators in Eq. (\ref{TB-GHam}). This operator exchange effectively inverts the conduction and valence bands for this valley. This valley-dependent operator ordering reflects the time-reversal symmetry relation between the Dirac points. In summary, the low-energy effective Hamiltonian for the pristine graphene can be expressed compactly by introducing the valley index $\chi = \pm 1$ 
\begin{equation}
    \hat{\mathcal{H}}_{0\chi} (\bm{p}) = \chi  v_F \hat{\bm{\sigma}}\cdot \bm{p},
    \label{HamilGF_Prist}
\end{equation}
where $\hat{\bm{\sigma}}$ is the Pauli vector acting on the sub-lattice degree of freedom, and $\bm{p} = \hbar \bm{q}$ represents the momentum measured relative to the Dirac point. Solution of the eigenvalue problem yields the eigenenergies
\begin{equation}
    \mathscr{E}_{s \chi} (\bm{q}) = s \chi \hbar v_F q,
    \label{Eq:PGraphH_Energy}
\end{equation}
where $s$ is the band index. The band inversion discussed above sets $s\chi=1$ as the conduction band.

\subsection{Strained graphene and the Dirac oscillator}

Mechanical strain induces non-uniform atomic displacements that distort the honeycomb lattice geometry. These structural modifications perturb the interatomic coupling strengths, leading to strain-dependent hopping energies that can be expressed as $t_j = t + \Delta t_j$, where $t$ is the unperturbed hopping parameter of pristine graphene and $\Delta t_j\ll t$ represents the strain-induced variation at each nearest-neighbor bond. Under these hopping modifications, we derive the low-energy Hamiltonian near the Dirac points $\bm{K}_D^{\pm}$ incorporating strain effects. Substituting the perturbed hopping parameters into (\ref{f_k-general}) and neglecting second-order terms ($\Delta t_j q_i$), we obtain the pristine graphene matrix element from Eq. (\ref{f_k-lowE}) plus a strain-induced term
\begin{equation}
    f^{(1)}_{\bm{q}} = -\left( \Delta t_1  + \Delta t_2 e^{\pm i \frac{2\pi}{3} }  + \Delta t_3 e^{\mp i \frac{2\pi}{3} } \right).
\end{equation}
While higher-order approximations in the strain expansion can lead to additional physical phenomena (see Ref. \cite{P_Ramezani_2013}), we focus here on the dominant first-order effects. To linear order, the strained graphene Hamiltonian takes the form
\begin{equation}
    \hat{\mathcal{H}}_{\chi} (\bm{p}) = \chi v_F \hat{\bm{\sigma}}\cdot \left( \bm{p} + \chi  e \bm{A}_{5} \right),
    \label{Hamilt_StrainGraph}
\end{equation}
where the strain-induced pseudogauge field is given by
\begin{equation}
    \bm{A}_{5} = \frac{1}{2 e v_F}\left( \Delta t_2 + \Delta t_3 - 2 \Delta t_1 \, , \, \sqrt{3} \left( \Delta t_2 - \Delta t_3 \right) \right) .
    \label{pseudoGF-Def}
\end{equation}
This field couples to the momentum in a manner analogous to a magnetic vector potential, but with opposite signs in each valley. The resulting valley-dependent coupling motivates the interpretation of $\bm{A}_{5}$ as a pseudogauge field.  The subscript $5$ reflects its origin in the $\gamma^5$ matrix of relativistic Dirac theory when the Hamiltonian is formulated in a $(3+1)$-dimensional bispinor representation.

When atomic displacements are small compared to the lattice constant $a$, the hopping energy variations can be expressed to first order as  
\begin{equation}
    \Delta t_j = - \frac{\beta t}{a^2} \bm{\rho}_j \cdot ( \bm{u}_j -\bm{u}_0 ),
    \label{Hoping_mod}
\end{equation}
where $\bm{u}_0$ denotes the displacement of the central atom (blue atom in sub-lattice A, Fig.~\ref{fig:Graphene_Lattice}), $\bm{u}_j$ represents the displacements of its three nearest neighbors in sub-lattice B, and $\beta = -\partial \ln t / \partial \ln a \sim 2$ is the electron Gr\"uneisen parameter. The quantity $\bm{u}_j -\bm{u}_0$ corresponds to the relative displacement between an atom in sub-lattice A and its neighbors in sub-lattice B. Eq.~\eqref{Hoping_mod} indicates that the hopping modification scales with the projection of this relative displacement onto the unstrained bond vector $\rho_j$, effectively coupling the strain field to the electronic structure.

In the continuum limit, we introduce the displacement field $\bm{u}(\bm{r})$, allowing the relative displacement to be approximated as $\bm{u}_j -\bm{u}_0 \sim ( \bm{\rho}_j \cdot \nabla  ) \bm{u}$. Substituting this into Eq.~\eqref{Hoping_mod} and defining the linearized strain tensor 
\begin{equation}
    u_{mn} = \frac{1}{2} \left( \frac{\partial u_m}{\partial x_n} + \frac{\partial u_n}{\partial x_m} \right),
\end{equation}
we obtain the pseudogauge field \eqref{pseudoGF-Def} in terms of strain components:
\begin{equation}
    \bm{A}_{5} = -\frac{3 \beta t}{4 e v_F} \big( u_{11} - u_{22} \, , \, - 2 u_{12} \big).
    \label{pseudoGF-disp}
\end{equation}
In addition to $\bm{A}_{5}$, strain generates the so-called deformation potential \cite{B_Gorbar2021,P_Suzuura_2002}, characterized by $V_s \propto u_{11} + u_{22}$ which appears as a diagonal term in the low-energy Hamiltonian. This potential couples equally to both valleys and modifies the energy spectrum uniformly.

We now derive the conditions on the displacement field $\bm{u}(\bm{r})$ required to simultaneously produce (i) a constant out-of-plane pseudomagnetic field $\bm{B}_{5} = \nabla \times \bm{A}_{5}$ while maintaining (ii) a vanishing deformation potential ($V_s=0$). To satisfy these conditions, the simplest ansatz consists of a linear pseudo gauge field with $A_{5\,x}\propto y$ and $A_{5\,y}\propto x$, and a strain configuration enforcing $u_{22} = - u_{11}$ so the deformation potential $V_s$ vanishes. The displacement field satisfying this conditions is given by 
\begin{equation}
    \bm{u} = -\frac{c_{B}}{4}\left( 2 xy \, , \,   x^2 -  y^2\right),
\end{equation}
where $c_{B}$ quantifies the strain magnitude. Such a configuration naturally emerges in deformed graphene nanobubbles \cite{S_Levy_2010,G_de_2013}, where the 
symmetric lattice distortion preserves the $V_s$ condition while maintaining a homogeneous $\bm{B}_{5}$. It must be emphasized that this displacement field is oriented along the graphene's crystal lattice, in line with the one discussed in Ref. \cite{E_Guinea_2010}. The resulting displacement field generates a pseudogauge field of the form 
\begin{equation}
    \bm{A}_{5} = \frac{B}{2}(y,-x),
    \label{pseudoGF_Sym}
\end{equation}
where the field strength $B= \frac{3 \beta t c_B}{2 e v_F}$ combines material parameters ($\beta$, $t$, $v_F$) with the strain magnitude $c_B$. The corresponding pseudomagnetic field $\bm{B}^{5} = -B \hat{z}$ exhibits a strain-dependent orientation: an outward (tensile) strain produces a field along $-\hat{z}$, while inward (compressing) strain would reverse the direction to $+\hat{z}$.

With the pseudo gauge field $\bm{A}_{5}$ from Eq.~(\ref{pseudoGF_Sym}) we derive the operator identity
\begin{equation}
    \hat{\bm{\sigma}}\cdot \bm{A}_{5} \equiv - i \frac{B}{2} \hat{\bm{\sigma}}\cdot \left( \hat{\sigma}_z \bm{r} \right),
\end{equation}
where $\bm{r}$ is the in-plane position vector. Substituting this into the strained Hamiltonian (\ref{Hamilt_StrainGraph}) and defining the \textit{effective valley mass} $m_\chi  \equiv \chi \hbar |\bm{K}_{D}^{\chi}| / v_F$ and the \textit{characteristic frequency} $\omega = eB / 2 |m_\chi|$, it simplifies to
\begin{equation}
    \hat{\mathcal{H}}_{\chi} (\bm{p}) = \chi v_F \hat{\bm{\sigma}}\cdot \left( \bm{p} - i m_\chi \omega \hat{\sigma}_z \bm{r} \right). 
    \label{Hamilt_MS-Osc}
\end{equation}
The resulting Hamiltonian corresponds to a massless Moshinsky-Szczepaniak oscillator \cite{T_Moshinsky_1989} in two dimensions, expressed in the pseudospin basis of Pauli matrices \cite{Rao_2004},  but with the Fermi velocity $v_F$ replacing the speed of light $c$. In the Hamiltonian (\ref{Hamilt_MS-Osc}), the frequency $\omega$ originates from the strain-electron coupling rather than an external potential, emerging via the non-minimal coupling to the pseudogauge field $\bm{A}_{5}$. Crucially, its valley-dependent sign generalizes the original Moshinsky-Szczepaniak model, introducing a chirality-specific response to mechanical deformations. While the Moshinsky-Szczepaniak model has been explored in graphene systems (such as in monolayer/bilayer studies \cite{quimbay2013graphenephysicsdiracoscillator} or quantum dots with hybrid magnetic-Dirac oscillator couplings \cite{BELOUAD2016773}) our work reveals a fundamental distinction: here, the oscillator dynamics emerge naturally from the strain-induced pseudogauge field $\bm{A}_{5}$, without external potentials or magnetic fields. This intrinsic coupling directly links mechanical deformations to valley-polarized oscillatory states. This mechanism suggests a route to engineer valley-selective confinement purely through strain.

\subsection{Pseudo-Landau levels}

A hallmark of graphene’s Dirac physics is its non-equidistant Landau level spectrum under real magnetic fields \cite{FoaTorres_Roche_Charlier_2020,E_O_2011}. Remarkably, strain-induced pseudogauge fields produce analogous Landau levels, as confirmed experimentally via STM in nanobubbles \cite{S_Levy_2010}. To describe these levels, we introduce the valley-dependent generalized momentum $\hat{\bm{\Pi}}^{\chi} = \hat{\bm{p}} + \chi e \bm{A}_{5}$, and define the magnetic length $l_{B} = \sqrt{\hbar / eB}$ (assuming tensile strain, so that $B>0$). Crucially, the commutator $[\Pi^{\chi}_{x},\Pi^{\chi}_{y}] = i\chi \hbar^2 /l_{B}^2$ reveals a valley-sensitive algebra, leading to the chiral ladder operators
\begin{equation}
    \begin{array}{c}
         \hat{a}_{\chi} \equiv \frac{l_B}{\sqrt{2}\hbar} \left( \Pi^{\chi}_{x} + i\chi \Pi^{\chi}_{y} \right),  \\[6pt]
         \hat{a}_{\chi}^{\dagger} \equiv \frac{l_B}{\sqrt{2}\hbar} \left( \Pi^{\chi}_{x} - i\chi \Pi^{\chi}_{y} \right).
         \label{eq:ladder_ops}
    \end{array}
\end{equation}
Unlike real magnetic fields, these operators depend explicitly on the valley index $\chi$, enabling valley-asymmetric Landau level occupation.

Using the ladder operators defined in Eq.~\eqref{eq:ladder_ops}, the Hamiltonian \eqref{Hamilt_StrainGraph} takes distinct matrix forms for each valley:
\begin{align}
    \hat{\mathcal{H}}_{+} (\bm{p}) = \frac{\sqrt{2} \hbar v_F}{l_{B}}  \begin{pmatrix}
        0 & \hat{a}_{+}^{\dagger} \\
        \hat{a}_{+} & 0
    \end{pmatrix},
    \label{H_K_plus}
\end{align}
for the valley $\bm{K}_{D}^{+}$, and 
\begin{align}
    \hat{\mathcal{H}}_{-} (\bm{p}) = -\frac{\sqrt{2} \hbar v_F}{l_{B}}  \begin{pmatrix}
        0 & \hat{a}_{-} \\
        \hat{a}_{-}^{\dagger} & 0
    \end{pmatrix},
    \label{H_K_minus}
\end{align}
for the valley $\bm{K}_{D}^{-}$. The sign reversal and operator exchange between valleys arise from the nature of the pseudogauge field $\bm{A}_{5}$. This leads to valley-polarized Landau levels with identical energies but opposite chirality. Diagonalizing \eqref{H_K_plus} and \eqref{H_K_minus} yields the non-equidistant spectrum $\mathscr{E}_{n}^{\chi,s} = s \chi v_{F} \sqrt{2n e\hbar B}$ ($n= 0,1,2,\dots$), which is formally identical to that of graphene under a real magnetic field. However, due to the valley dependence in the Hamiltonian, the corresponding eigenstates exhibit a striking valley-dependence. For $\chi = 1$:
\begin{align}
    \psi^{+,s}_{n} =\frac{1}{\sqrt{2}} \begin{pmatrix}
        \ket{n} \\
        s \ket{n-1}
    \end{pmatrix},
    \label{eig_plus}
\end{align}
while for $\chi = -1$: 
\begin{align}
    \psi^{-,s}_{n} =\frac{1}{\sqrt{2}} \begin{pmatrix}
        - s \ket{n-1} \\
        \ket{n}
    \end{pmatrix},
    \label{eig_minus}
\end{align}
where $\ket{n}$  denotes the $n$-th Landau level state. In contrast to real magnetic fields, where the $n=0$ Landau level is valley-degenerate and appears in both conduction and valence bands, the strain-induced pseudogauge field produces a fundamentally different configuration. As evident from Eqs. \eqref{eig_plus} and \eqref{eig_minus}, the zero-energy state becomes valley-polarized, localizing exclusively in the conduction band (for the tensile strain considered). This broken valley degeneracy enables novel transport phenomena: while real magnetic fields generate counter-propagating valley currents that cancels out, strain-engineered pseudofields permit net valley currents, a unique feature with potential applications in valleytronics.

\section{Kinetic theory approach to thermoelectric transport} \label{Transport_Section}

In typical graphene samples, strain profiles are confined to nanoscale regions on the order of $2$-$10$ nm \cite{S_Levy_2010}. When considering macroscopic graphene sheets, these nano-deformations can be effectively modeled as randomly distributed localized perturbations. Following the approach established in Section \ref{Model}, we represent such strain-induced defects as regions of effective pseudomagnetic fields. While the assumption of perfectly constant $B_5$ within each random impurity region is an idealization, it is discussed in Ref. \cite{E_Guinea_2010} that the pseudomagnetic field is predominantly constant over most of the deformation's central area. This approximation proves sufficient for studying electronic transport within the Boltzmann regime. To investigate how these strain-induced impurities affect electronic transport, we compute the thermoelectric coefficients (including the DC conductivity, thermal conductivity, and Seebeck coefficient) using the Boltzmann transport equation within the relaxation time approximation.

\subsection{Transport relaxation time}
The transport relaxation time $\tau_{\text{tr}}(k)$ is determined by impurity scattering processes through the integral relation \cite{ziman,ziman_principles,mahan}
\begin{equation}
    \frac{1}{\tau_{\text{tr}}(\bm{k})}=n_i\int\frac{d^2k'}{(2\pi)^2} W_{\bm{k'}\bm{k}}(1-\cos\theta_{\bm{k'}}),\label{eq:tau_1}
\end{equation}
where $n_i$ denotes the 2D impurity density, $\theta_{\bm{k'k}}$ is the scattering angle between initial ($\bm{k}$) and final ($\bm{k'}$) states, and $W_{\bm{k'}\bm{k}}$ represents the quantum transition rate. While $n_i$ is typically considered an experimentally tunable parameter, nanobubble formation can be thermally-induced, leading to a temperature-dependent modulation of the impurity density \cite{PhysRevB.106.045418}. For elastic scattering processes, Lippmann-Schwinger equation yields
\begin{equation}
    W_{\bm{k'}\bm{k}} =\frac{2\pi }{\hbar}|T_{\bm{k}'\bm{k}}|^2\,\delta(\mathscr{E}_{\bm{k}}-\mathscr{E}_{\bm{k'}}),
    \label{eq:golden_rule}
\end{equation}
where $T_{\bm{k'}\bm{k}}$ are the matrix elements of the transition operator \cite{Zettili2009,delaPena2014}. While the first Born approximation leads to Fermi's Golden Rule, an exact expression for $T_{\bm{k'}\bm{k}}$ can be obtained by solving the scattering problem through partial waves analysis. To this end, we exploit the Dirac oscillator structure of the strained graphene Hamiltonian in Eq.~\eqref{Hamilt_MS-Osc}. Crucially, the total angular momentum operator $\hat{\mathcal{J}}_z = \hat{\mathcal{L}}_z \otimes \hat{\sigma}_0 + \hat{\mathcal{S}}_z$ commutes with $\hat{\mathcal{H}}_{\chi}(\bm{p})$. This conservation law enforces solutions with definite total angular momentum $m_j = \pm\frac{1}{2}, \pm\frac{3}{2},...$, expressed in polar coordinates as 
\begin{equation}
    \Psi_{\chi}^{m_j} = \begin{pmatrix}
        f^{\chi}_{m_j-1/2}(r) e^{i(m_j-1/2)\theta} \\[6pt]
        i s g^{\chi}_{m_j+1/2}(r) e^{i(m_j+1/2)\theta}
    \end{pmatrix}.
\end{equation}
Substituting into the time-independent Dirac equation $ \hat{\mathcal{H}}_{\chi}(\bm{p}) \Psi_{\chi}^{m_j} = \mathscr{E} \Psi_{\chi}^{m_j} $, we obtain the coupled system:
\begin{align}
    &\left[ -\frac{1}{r}\partial_r(r\partial_r) + \frac{(m_j\mp1/2)^2}{r^2} + k_\chi^2 r^2 \right] \begin{pmatrix} f^{\chi} \\ g^{\chi} \end{pmatrix} \notag\\
    &\quad = \left[ \left(\frac{\mathscr{E}}{\hbar v_F}\right)^2 + 2 k_\chi (m_j \pm 1/2) \right] \begin{pmatrix} f^{\chi} \\ g^{\chi} \end{pmatrix},
    \label{DiracEq_spinors}
\end{align}
where $k_\chi \equiv m_{\chi}\omega/\hbar$. Since we model the pseudomagnetic field as a perturbation confined to $r<R$ (with $R \sim 2$-$10$ nm), the electronic states outside this region ($r \geq R$) are governed by the pristine graphene Hamiltonian in Eq.~\eqref{HamilGF_Prist}. Consequently, the allowed energies are given by Eq.~\eqref{Eq:PGraphH_Energy}, i.e., $\mathscr{E} = \mathscr{E}_{s\chi}(\bm{k})$. The scattering problem is solved by enforcing continuity of the solutions across the boundary $r=R$:\\
\noindent
\textit{Inner solution} ($r<R$): Exact oscillator solutions to Eq.~\eqref{DiracEq_spinors}:
    \begin{align}
\Psi_{\chi}^{m_j} =&  C \exp \left(- \frac{1}{2} k_\chi r^2 \right) r^{m_{j}-1/2} \notag\\
&\,\, \times\begin{pmatrix}  L_{\scriptstyle{ \nu_{+} }} ^{\scriptstyle{m_{j} - 1/2}} \left( k_\chi r^2 \right) \,  e ^{i (m_{j} -1/2) \theta } \\[6pt]
    i s \frac{2 k_\chi}{k}\, r\, L_{\scriptstyle{ \nu_{-} }} ^{\scriptstyle{m_{j} + 1/2}} \left( k_\chi r^2 \right) \,  e ^{i (m_{j} +1/2) \theta }  \end{pmatrix},
    \label{DiracEq_Inner}
    \end{align}
where $L_\nu^a(x)$ are generalized Laguerre functions and $\nu_{\pm} = k^{2} / 4k_\chi - (1 \mp 1)/2$.\\
\noindent
\textit{Outer solution}($r\geq R$): Pristine graphene waves:
    \begin{equation}
\Phi_{\bm{k}}^{m_j}\!\! = \!\!\begin{pmatrix}
            \left[ c _{1} \, J _{m_{j} -1/2} (kr) + c _{2} \, Y _{m_{j} -1/2} (kr)  \right]\! e ^{i (m_{j} -1/2) \theta } \\[6pt] \! i s  \!\left[ c _{1} \, J _{m_{j} +1/2} (kr) + c _{2} \, Y _{m_{j} +1/2} (kr) \right] \!  e ^{i (m_{j} +1/2) \theta }
        \end{pmatrix},
    \end{equation}
    where $J_\nu$, $Y_\nu$ are Bessel functions of the first and second kind.
The phase shifts $\delta_{m_j}^{\chi}$, encoding all scattering information, are given by
\begin{align}
    &\tan \delta_{m_j}^{\chi} \notag\\
    &=  \frac{ J _{m_{j} -1/2} (kR) \, g^{\chi}_{m_j+1/2}(r) - J _{m_{j} +1/2} (kR) \, f^{\chi}_{m_j-1/2}(r)}{ Y _{m_{j} -1/2} (kR) \, g^{\chi}_{m_j+1/2}(r) - Y _{m_{j} +1/2} (kR) \, f^{\chi}_{m_j-1/2}(r) },\label{eq:delta_mj-NE}
\end{align}
or, written explicitly in terms of the spinor in Eq.~\eqref{DiracEq_Inner}:
\begin{widetext}
\begin{equation}
    \tan \delta_{m_j}^{\chi} =  \frac{ 2  k_\chi R \, J _{m_{j} -1/2} (kR) \, L_{\scriptstyle{ \nu_{-} }} ^{\scriptstyle{m_{j} + 1/2}}\left( k_\chi R^2\right) - k J _{m_{j} +1/2} (kR) \, L_{\scriptstyle{ \nu_{+}}} ^{\scriptstyle{m_{j} - 1/2}} \left(  k_\chi R^2 \right) }{ 2  k_\chi R \, Y _{m_{j} -1/2} (kR) \, L_{\scriptstyle{ \nu_{-}}} ^{\scriptstyle{m_{j} + 1/2}}\left( k_\chi R^2\right) - k Y _{m_{j} +1/2} (kR) \, L_{\scriptstyle{ \nu_{+}}} ^{\scriptstyle{m_{j} - 1/2}} \left(  k_\chi R^2 \right) }.\label{eq:delta_mj}
\end{equation}
\end{widetext}
These phase shifts generate valley-asymmetric scattering through the $\chi$-dependence of $k_\chi$. It is worth mentioning that phase shifts obtained through a similar methodology have been used to study transport and other effects in Weyl semimetals \cite{Bonilla_1,Bonilla_2,Bonilla_3,PhysRevResearch.2.012043}.

The transition matrix $T$ connects to measurable quantities through the differential cross section \cite{Zettili2009, delaPena2014}. For graphene's massless Dirac fermions, the asymptotic behavior of scattered waves yields 
\begin{equation}
    \frac{d\sigma}{d\theta} =\frac{k}{2\pi\hbar^2v_F^2}|T_{\bm{k'}\bm{k}}|^2.\label{eq:cross_section_T}
\end{equation}
Calculating the cross section in terms of the phase shifts \cite{S_Ramezani_2011}, the square modulus of $T$ is written as
\begin{equation}
    |T_{\bm{k'}\bm{k}}|^2 \!=\!\frac{4\hbar^2 v_F^2}{k^2}\!\!\sum_{m_j,m'_j}\!\!e^{i(m_j-m'_j)\theta}e^{i(\delta_{m_j}^{\chi}-\delta_{m'_j}^{\chi})}\!\sin\delta_{m_j}^{\chi}\!\sin\delta_{m'_j}^{\chi}.\label{eq:T_phase_shifts}
\end{equation}
The transport relaxation time integral \eqref{eq:tau_1} can be evaluated by substitution of the quantum transition rate \eqref{eq:golden_rule} in terms of Eq.~\eqref{eq:T_phase_shifts}. Integration requires specifying the carrier type (electrons/holes) through the energy sign $E_{\bm{k}}$. Crucially, while this fixes the band index $s$, the valley dependence remains as the phase shifts are band independent. Direct integration and further manipulation of trigonometrical functions yields
\begin{equation}
    \frac{1}{\tau_{\text{tr}}(\bm{k})}= \frac{ 2n_i v_F}{ k} \sum_{m_j=-\infty}^{\infty}\sin^2\left(\delta_{m_j}^{\chi}-\delta_{m_j-1}^{\chi}\right).\label{eq:tau_final_2}
\end{equation}
At first glance, the relaxation time in Eq.~\eqref{eq:tau_final_2} appears to depend on the valley index $\chi$. However, as evident from eqs. \eqref{DiracEq_spinors}, the system exhibits a symmetry under simultaneous sign reversal of $m_j$ and $k_{\chi}$: transforming $m_j \to -m_j$ and $k_{\chi} \to k_{-\chi}=-k_{\chi}$, in either the upper or lower component yields the other equation. This implies the relation $g^{\chi}_{\alpha} = f^{-\chi}_{-\alpha}$. By applying this sign transformation to Eq.~\eqref{eq:delta_mj-NE} and exploiting the properties of Bessel functions $J_{-\alpha}(x) = (-1)^{\alpha} J_{\alpha}(x)$ and $Y_{-\alpha}(x) = (-1)^{\alpha} Y_{\alpha}(x)$ \cite{abramowitz1965handbook}, we demonstrate that the phase shifts satisfy $\delta_{m_j}^{\chi} = \delta_{-m_j}^{-\chi}$. This relation is illustrated in Fig.~\ref{fig:delta_vs_mj}. Consequently, the summation in Eq.~\eqref{eq:tau_final_2} becomes identical for both valleys, rendering the relaxation time valley-independent. This behavior is further confirmed by the numerical results shown in Fig.~\ref{fig:tau_vs_EF}.

\subsection{Transport coefficients}

Within the relaxation time approximation, the distribution function $f_{\sigma s\chi}(\bm{k})$ (for each spin $\sigma$, band $s$ and valley $\chi$ channel) evolves as \cite{ziman}
\begin{equation}
	\dot{\bm{x}} \cdot \nabla_{\bm{x}} f_{\sigma s\chi} + \dot{\bm{k}} \cdot \nabla_{\bm{k}} f_{\sigma s\chi} = -\frac{g_{\sigma s\chi}}{\tau_{\text{tr}}(\bm{k},\chi)},
\label{eq:Boltz_1}
\end{equation}
where $g_{\sigma s\chi} \equiv f_{\sigma s\chi} - f^0_{\sigma s\chi}$ measures the deviation from local equilibrium, described by the Fermi-Dirac distribution
\begin{equation}
f^0_{\sigma s\chi}(\bm{k}) = \left[1 + e^{(\mathscr{E}_{s\chi}(\bm{k}) - \mu(\bm{x}))/k_B T(\bm{x})}\right]^{-1},
\label{eq:f_0_local}
\end{equation}
where $T(\bm{x})$ is the local equilibrium temperature, $\mu(\bm{x})$ is the local chemical potential and $k_B$ is the Boltzmann constant. The semiclassical dynamics (neglecting Berry curvature) is governed by
\begin{equation}
\dot{\bm{x}} = \frac{1}{\hbar} \nabla_{\bm{k}} \mathscr{E}_{s\chi}(\bm{k}), \quad
\dot{\bm{k}} = \frac{q}{\hbar} (\bm{E} + \dot{\bm{x}} \times \bm{B}),
\end{equation}
where $q$ is the carrier charge. For $\bm{B}=0$ and weak perturbations, the linearized solution becomes:
\begin{equation}
g_{\sigma s\chi} \!=\! -\frac{\tau_{\text{tr}}(\bm{k},\chi)}{\hbar} \!\left( \frac{\partial f^0}{\partial T} \nabla T + \frac{\partial f^0}{\partial \mu} \nabla \mu + q \frac{\partial f^0}{\partial \mathscr{E}} \bm{E} \right) \cdot \nabla_{\bm{k}} \mathscr{E}_{s\chi},
\label{eq:Boltz_4}
\end{equation}
where we suppressed indices on $f^0$ for clarity. This form explicitly shows the coupling between thermal gradients ($\nabla T$), chemical potential gradients ($\nabla \mu$) and electric fields ($\bm{E}$), through the transport relaxation time $\tau_{\text{tr}}(\bm{k},\chi)$ and band structure $\mathscr{E}_{s\chi}(\bm{k})$. Substitution of the equilibrium distribution \eqref{eq:f_0_local} into \eqref{eq:Boltz_4} yields the equilibrium deviation in terms of thermodynamic forces
\begin{align}
g_{\sigma s\chi} &= \tau_{\text{tr}}(\bm{k},\chi) \left(-\frac{\partial f^0}{\partial \mathscr{E}}\right) \bm{v}_{s\chi}(\bm{k}) \notag\\
&\,\,\times\left[ q\left(\bm{E} - \frac{1}{q}\nabla\mu\right) + \left(\frac{\mathscr{E}_{s\chi} - \mu}{T}\right)(-\nabla T) \right],
\label{eq:Boltz_5}
\end{align}
where $\bm{v}_{s\chi}(\bm{k}) = \hbar ^{-1} \nabla_{\bm{k}} \mathscr{E}_{s\chi}$ is the group velocity and the terms in brackets represent the electrochemical field and the thermal driving force respectively.  

Following Eq.~\eqref{eq:Boltz_5}, the total electric DC current ($\bm{J}^{1}$) and the heat flow ($\bm{J}^{2}$) are written as
\begin{equation}
    \bm{J}^1=\frac{q^2}{T}\bm{L}^{(11)}\left(\bm{E}-\frac{1}{q}\nabla \mu\right)+\frac{q}{T^2}\bm{L}^{(12)}(-\nabla T),
    \label{eq:J1}
\end{equation}
\begin{equation}
    \bm{J}_2=\frac{q}{T}\bm{L}^{(21)}\left(\bm{E}-\frac{1}{q}\nabla \mu\right)+\frac{1}{T^2}\bm{L}^{(22)}(-\nabla T),
    \label{eq:J2}
\end{equation}
where the Onsager coefficients $\bm{L}^{(ij)}$, governing all electronic transport phenomena, are defined by the integral expression
\begin{align}
\bm{L}^{(ij)} =& T \sum_{\sigma,s,\chi} \int \frac{d^2k}{(2\pi)^2} \tau_{\text{tr}}(\bm{k},\chi)
\left[\mathscr{E}_{s\chi}(\bm{k}) - \mu\right]^{i+j-2}\notag\\
&
\quad \times\left(-\frac{\partial f^0}{\partial \mathscr{E}}\right)
\bm{v}_{s\chi}(\bm{k}) \otimes \bm{v}_{s\chi}(\bm{k}).
\label{eq:L_ij}
\end{align}
For graphene's Dirac fermions, with $\bm{v}_{s\chi}(\bm{k}) = s\chi v_{F} \hat{k}$, these tensors become diagonal. Owing the independence of the relaxation time on the spin and valley, sums on those index yields a factor of $4$. Also, due to the band selection trough the Fermi level, the sum on $s$ yields a factor of $1$. In the zero-temperature limit, the derivative $\, -\partial f^0 / \partial \mathscr{E} \to \delta (\mathscr{E}-\mathscr{E}_F) \, $, with $\mathscr{E}_F$ the Fermi level, simplifies the integration, yielding exact closed-form expressions. To study the thermoelectric behavior for low temperatures (i.e., $k_{B} T \ll \mathscr{E}_{F}$), we apply the Sommerfeld expansion to obtain the leading-order contributions \cite{ziman_principles}
\begin{align}
    L^{(11)}_{\alpha\beta}(T)=&\frac{T \mathscr{E}_F}{\pi \hbar^2}\delta_{\alpha\beta} \Big\lbrace \tau_{\text{tr}}(\mathscr{E}_F)\notag\\
    &\quad +\frac{(\pi k_B T)^2}{6 \mathscr{E}_F}\left[2\tau_{\text{tr}}^{\prime}(\mathscr{E}_F)+\mathscr{E}_F\tau_{\text{tr}}^{\prime \prime}(\mathscr{E}_F)\right]\Big\rbrace,\label{eq:L_11}\\
    L^{(12)}_{\alpha\beta}(T) =& L^{(21)}_{\alpha\beta}(T) \notag\\
    =& \frac{\pi k_B^2 T^3}{3 \hbar^2}\delta_{\alpha\beta} \left\lbrace\tau_{\text{tr}}(\mathscr{E}_F)+\mathscr{E}_F\tau'_{\text{tr}}(\mathscr{E}_F)\right\rbrace, 
    \label{eq:L_12}\\
    L^{(22)}_{\alpha\beta}(T) =& \frac{\pi k_B^2 T^3}{3 \hbar^2}\delta_{\alpha\beta} \mathscr{E}_F\,\tau_{\text{tr}}(\mathscr{E}_F).\label{eq:L_22}
\end{align}

The key transport-related quantities follow directly from the Onsager tensors. DC conductivity $\bm{\sigma}^{\text{\text{DC}}}$ under isothermal conditions is
\begin{equation}
    \bm{\sigma}^{\text{DC}} =\frac{q^2}{T}\bm{L}^{(11)},
    \label{eq:cond_DC}
\end{equation}
which after replacing the corresponding Onsager coefficient leads to
\begin{align}
\sigma_{\alpha\beta}(T)=&2\frac{q^2}{ h}\delta_{\alpha\beta}\left\lbrace  \frac{\mathscr{E}_F}{\hbar}\, \tau_{\text{tr}}(\mathscr{E}_F)\right.\notag\\
&\quad \left.+\frac{\pi^2}{6}\frac{(k_B T)^2}{\hbar} \left[2\tau'_{\text{tr}}(\mathscr{E}_F)+\mathscr{E}_F\tau''_{\text{tr}}(\mathscr{E}_F)\right]\right\rbrace,\label{eq:cond_DC_final}
\end{align}
where $h$ is the Planck constant. The residual conductivity at zero temperature is
\begin{align}
\sigma_{\alpha\beta}(T\ra 0)=2\frac{q^2}{ h} \delta_{\alpha\beta} \frac{\mathscr{E}_F}{\hbar}\, \tau_{\text{tr}}(\mathscr{E}_F).\label{eq:cond_DC_zero}
\end{align}
The electronic thermal conductivity $\bm{\kappa}^{el}$, obtained by setting $\bm{J}^{1} = 0$ in Eq.~\eqref{eq:J1}, solving for the electrochemical field and substituting in Eq. \eqref{eq:J2}, becomes
\begin{equation}
    \bm{\kappa}^{\text{el}} =\frac{1}{T^2}\left\lbrace \bm{L}^{(22)}-\bm{L}^{(12)}\left[\bm{L}^{(11)}\right]^{-1}\bm{L}^{(21)} \right\rbrace. \label{eq:cond_term}
\end{equation}
The first term in Eq.~\eqref{eq:cond_term} is of order $T$ whereas the second term is of order $T^3$. Then, to the lowest order in temperature, we approximate the thermal conductivity as
\begin{equation}
    \kappa^{\text{el}}_{\alpha\beta} \approx \frac{\pi k_B^2 T}{3 \hbar^2}\delta_{\alpha\beta} \mathscr{E}_F\,\tau_{\text{tr}}(\mathscr{E}_F).\label{eq:cond_term_final}
\end{equation}
Seebeck coefficient of thermopower, defined as $S=-\Delta V/\Delta T$ ($\Delta V$ being the difference of potential) is also written in terms of Onsager coefficients as
\begin{equation}
    \bm{S}=\frac{1}{qT}\left[\bm{L}^{(11)}\right]^{-1}\bm{L}^{(12)}.\label{eq:Seebeck}
\end{equation}

\section{Results and analysis}
\label{Results}

In this section, we perform the computation of the DC and thermal conductivities, as well as the thermoelectric analysis.  The existence of nanobubbles in graphene has been experimentally confirmed, such that the strain induces the appearance of giant pseudogauge fields ranging in magnitude from 200 to 600 Tesla (T) \cite{S_Levy_2010}. These fields are approximately constant and uniform in the direction perpendicular to the plane of the graphene sheet. The diameter of the nanobubbles can range from 4 to 10 nm, with a height of 0.3 to 2 nm \cite{S_Levy_2010}.  We obtain an estimate of the nanobubble density from Ref.~\cite{Lim2013} that explores the creation and properties of graphene nanobubbles on diamond surfaces. They report a nanobubble density of $n_b\sim 8\times10^{10}$ cm$^{-2}$. Additionally, we set the Fermi velocity to $v_F \sim 10^{6}$ m/s, in agreement with previously reported values for graphene \cite{DasSarma2011ElectronicTransport}, and consider electrons as the charge carriers. Thus, we select $s\chi = +1$ and assign the electronic charge as $q = -e = -1.602\times10^{-19}$ C. On the other hand, typical values of the Fermi energy in graphene are of the order $\mathscr{E}_F\sim 200$ meV, which corresponds to a density of carriers of approximately $n_c\sim 10^{12}$ cm$^{-2}$ \cite{DasSarma2011ElectronicTransport}. It is worth noting that this choice of parameters falls within the commonly accepted range of validity for the linear approximation in graphene, which remains accurate for energies up to approximately $1\,\text{eV}$, so that band-bending effects are negligible. The minimum conductivity measured in graphene is $\sigma_{xx}= 4 e^2/h\approx 115\, \mu$S, a value which takes into account the spin and valley symmetry \cite{Tan2007Measurement,Kim2016ValleySymmetry}.
\begin{figure}[!ht]
    \centering    \includegraphics[width=0.9\linewidth]{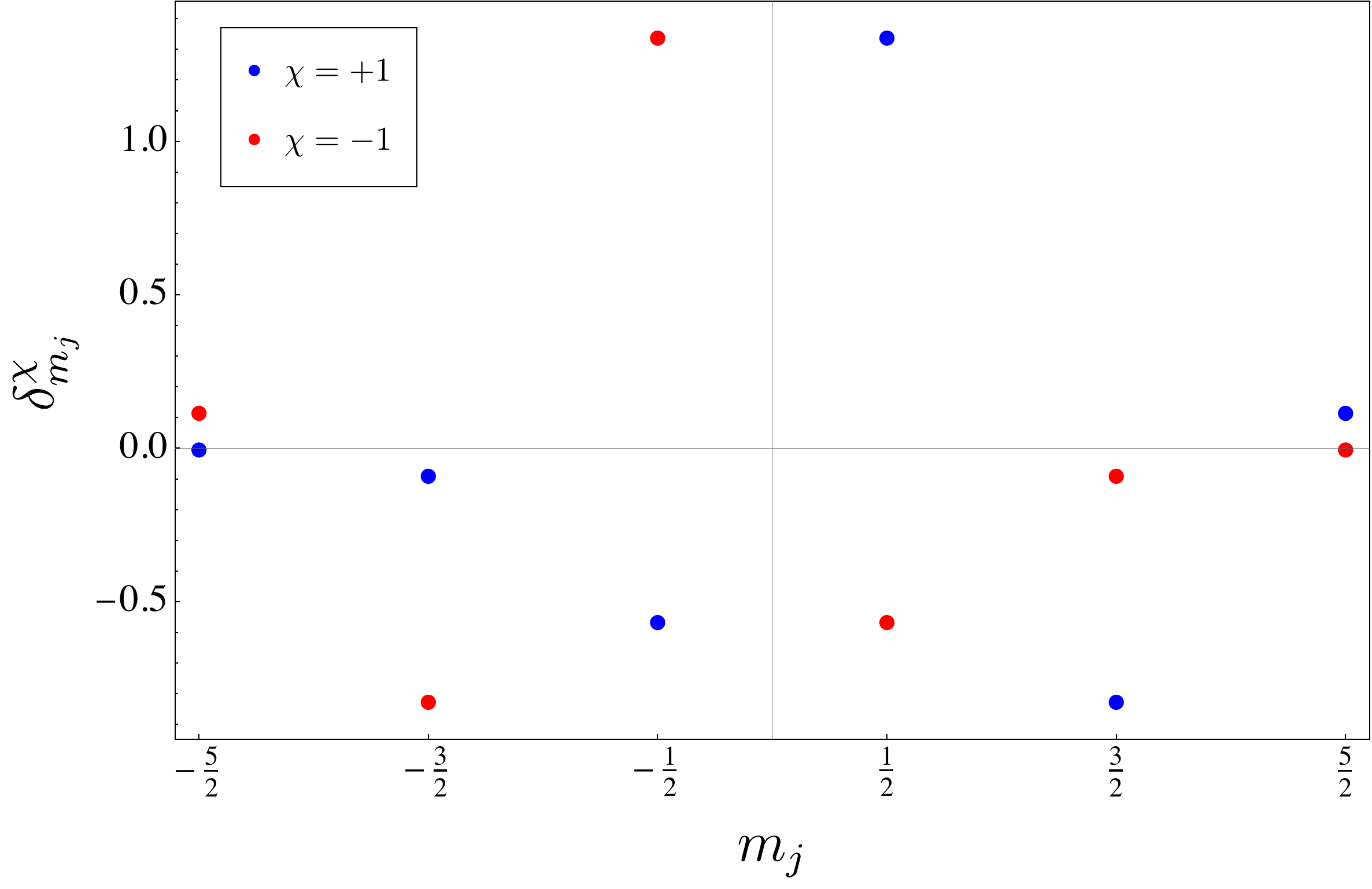}
    \caption{The phase shifts for each valley $\chi$, computed from Eq.~\eqref{eq:delta_mj} for a 4 nm-radius nanobubble under a constant 400 T pseudomagnetic field and a Fermi energy of 200 meV. } 
    \label{fig:delta_vs_mj}
\end{figure}

\begin{figure}[b]
    \centering
    \begin{subfigure}{0.45\textwidth}
         \centering
         \includegraphics[width=\linewidth]{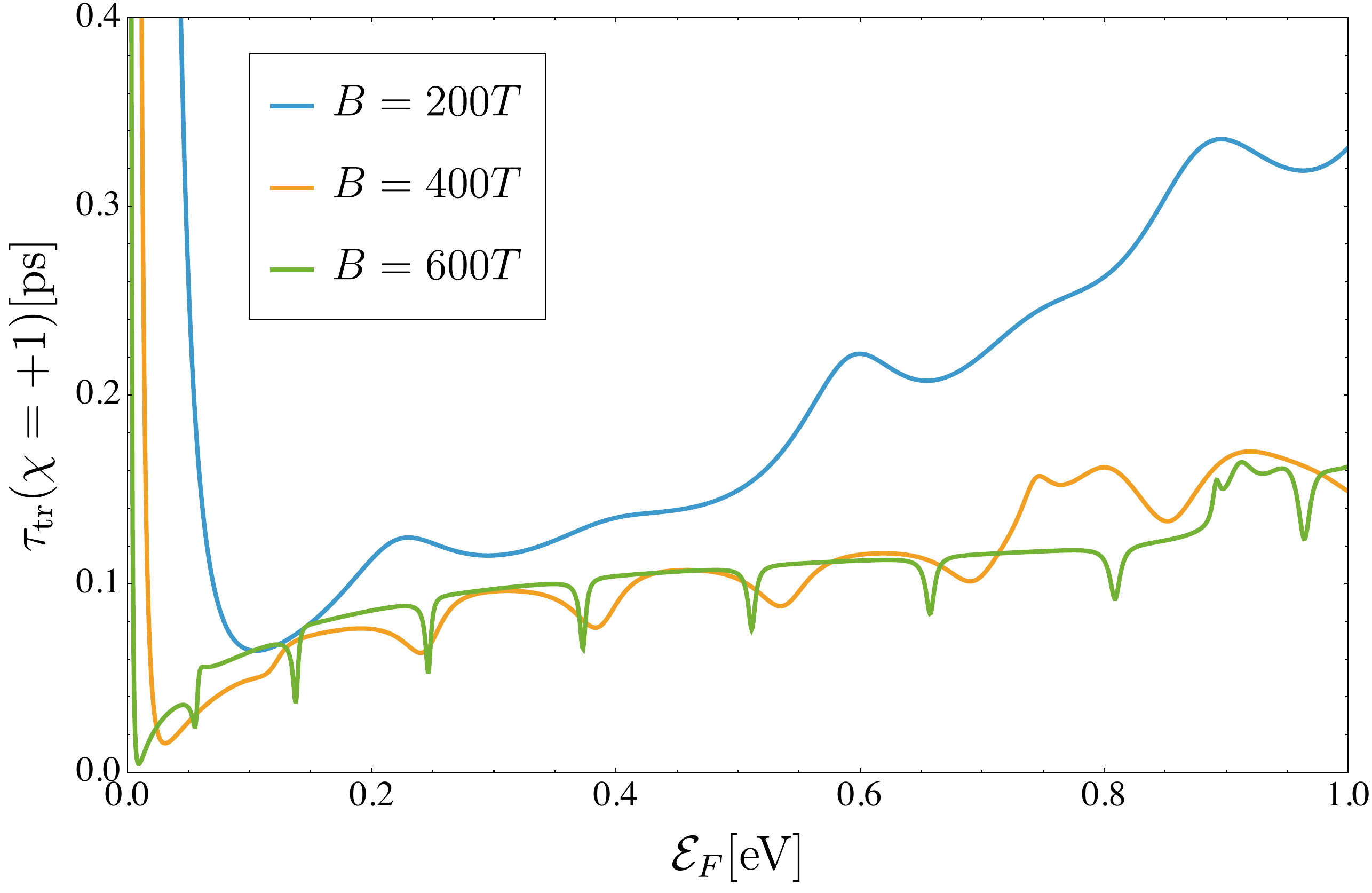}
         \subcaption{}\label{fig:tau_pos}
     \end{subfigure}
     \begin{subfigure}{0.45\textwidth}
         \centering
         \includegraphics[width=\linewidth]{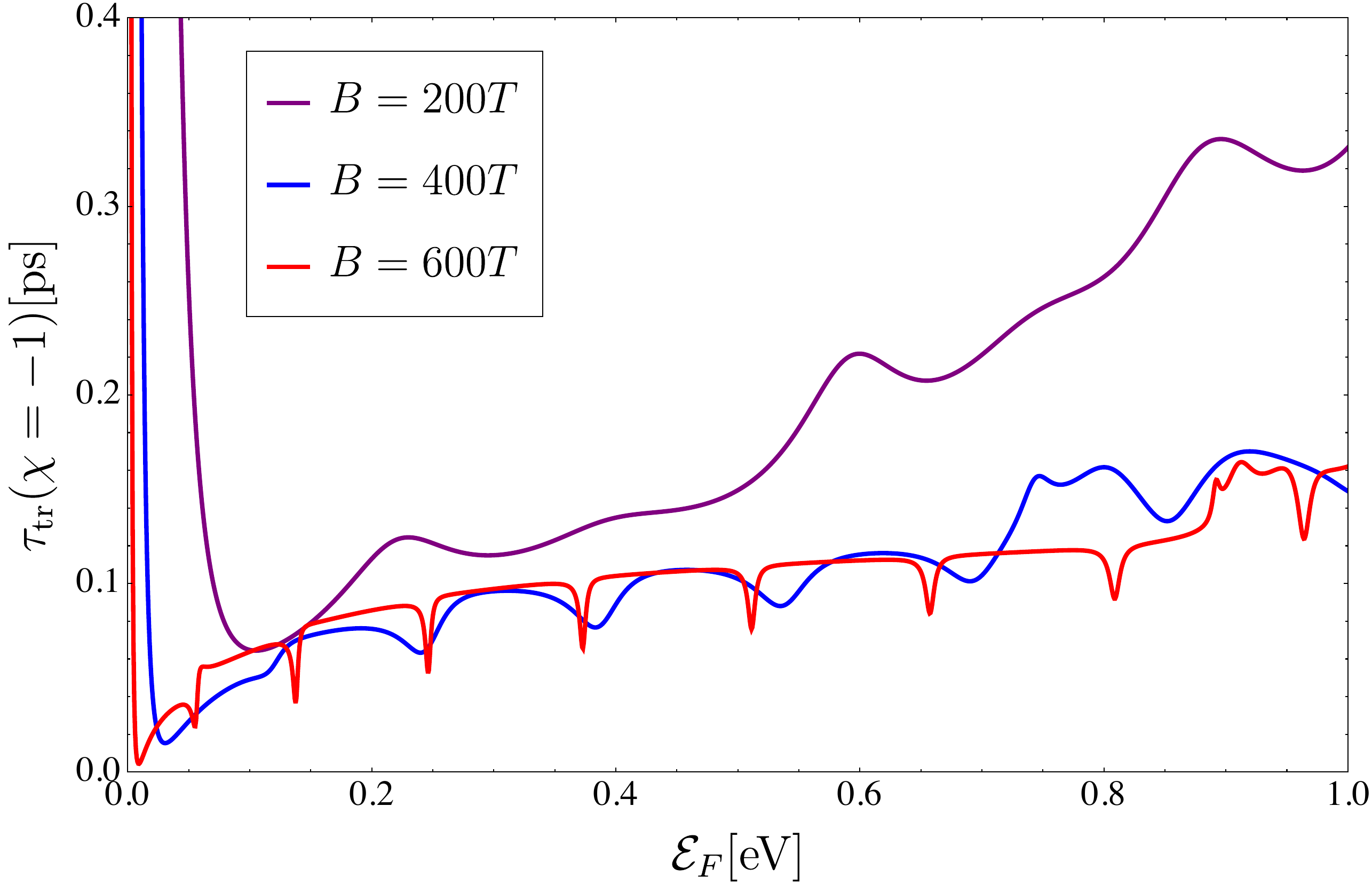}
         \subcaption{}\label{fig:tau_neg}
     \end{subfigure}
    \caption{Transport relaxation time (in picoseconds) as a function of Fermi energy, calculated using Eq.~\eqref{eq:tau_final_2} for a random distribution of 4 nm-radius nanobubbles with a concentration of $n_b = 8 \times 10^{10}$ cm$^{-2}$, and for various pseudomagnetic fields. The subfigure (a) corresponds to the $\chi=+1$ valley, whereas (b) corresponds to the $\chi=-1$ valley.} 
 \label{fig:tau_vs_EF}
\end{figure}

\paragraph{The role of the valley index $\chi$.} An important feature of strain-induced pseudogauge fields is their opposite coupling sign at each valley $\chi$. As discussed after Eq.~\eqref{eq:tau_final_2}, the phase shifts indeed depend on the valley sign $\chi$. However, the phase shifts $\delta_{m_j}$ for a given valley $\chi$ coincide with the phase shifts $\delta_{-m_j}$ of the opposite valley $-\chi$. This is clearly shown in Fig.~\ref{fig:delta_vs_mj}, where the phase shifts are computed for each valley and for different values of the total angular momentum $m_j$. Since the sum in Eq.~\eqref{eq:tau_final_2} for the transport relaxation time is symmetric with respect to $m_j$, the total contribution from each valley is the same. As a result, the relaxation time of the transport does not depend on the sign of the valley, as can be seen in Fig.~\ref{fig:tau_pos}-\ref{fig:tau_neg}. This outcome is expected, as changing the valley sign is equivalent to reversing the magnetic field in the original valley, and it is well-known that longitudinal transport depends only on the magnitude of the magnetic field. Similarly, a physical magnetic field that couples with the same sign to both valleys does not produce any difference in the transport relaxation time between valleys. Therefore, achieving a valley polarization effect requires the combination of a physical magnetic field with a torsion-induced pseudomagnetic field. In that case, the effective magnetic field, defined as the sum of both, differs in magnitude between the two valleys \cite{Munoz_2017}.

\paragraph{The transport relaxation time.} Figure \ref{fig:tau_vs_EF} shows the behavior of the relaxation time as a function of the Fermi energy, computed using Eq.~\eqref{eq:tau_final_2}, for each valley and for several values of the pseudomagnetic field. It is important to note that the relaxation time does not always increase monotonically with the Fermi energy, although it generally tends to grow as the energy increases. The minima that appear, becoming more pronounced as the magnitude of the pseudomagnetic field increases, are associated with the resonant scattering with the quasi-bound states inside the nanobubbles \cite{resonant_1,resonant_2,resonant_3,resonant_4}. In our semiclassical treatment of the Boltzmann equation, the transport relaxation time, $\tau_\text{tr}(\mathscr{E})$, is expressed in the $T$-matrix formalism  as a sum over partial-wave phase shifts $\delta_{m_j}(\mathscr{E})$, where each phase shift undergoes a sharp jump when crossing some resonant energy. This abrupt phase change induces destructive interference between the resonant contribution of the $m_j$-th channel of total angular momentum and the continuous background, leading to cancellation of $\sin^2\left(\delta_{m_j}-\delta_{m_j-1}\right)$ near the resonant energy, which appears as a pronounced dip in $\tau_\text{tr}(\mathscr{E})$. It is important to note that the effect of resonant scattering cannot be captured by expressing the inverse of the relaxation time using Fermi's golden rule within the first Born approximation, as is commonly done in the semiclassical Boltzmann theory. This approximation does not incorporate the analytic structure of the $S$-matrix, where bound states appear as complex poles that account for the resonances. However, these features are fully included in our $T$-matrix-based approach.

The quasi-bound states in our case would correspond to the pseudo-Landau levels that would be expected if the pseudomagnetic field extents uniformly throughout the sample. However, in this problem, the finite size of the nanobubbles plays an important role. When the radius of the nanobubble is taken into account, the quasi-bound states localized in the bubbles no longer match the usual pseudo-Landau levels \cite{Belouad_2020,ELAZAR2024112573,ELAZAR2024416005}. 

Another general trend is that the transport relaxation time decreases as the magnitude of the torsional pseudofield increases. A stronger field produces two effects: it generates more quasi-bound states within the bubbles, which enhances scattering, and it deflects carriers more effectively.
\begin{figure}[!ht]
\centering
    \begin{subfigure}[]{0.45\textwidth}
         \centering
         \includegraphics[width=\linewidth]{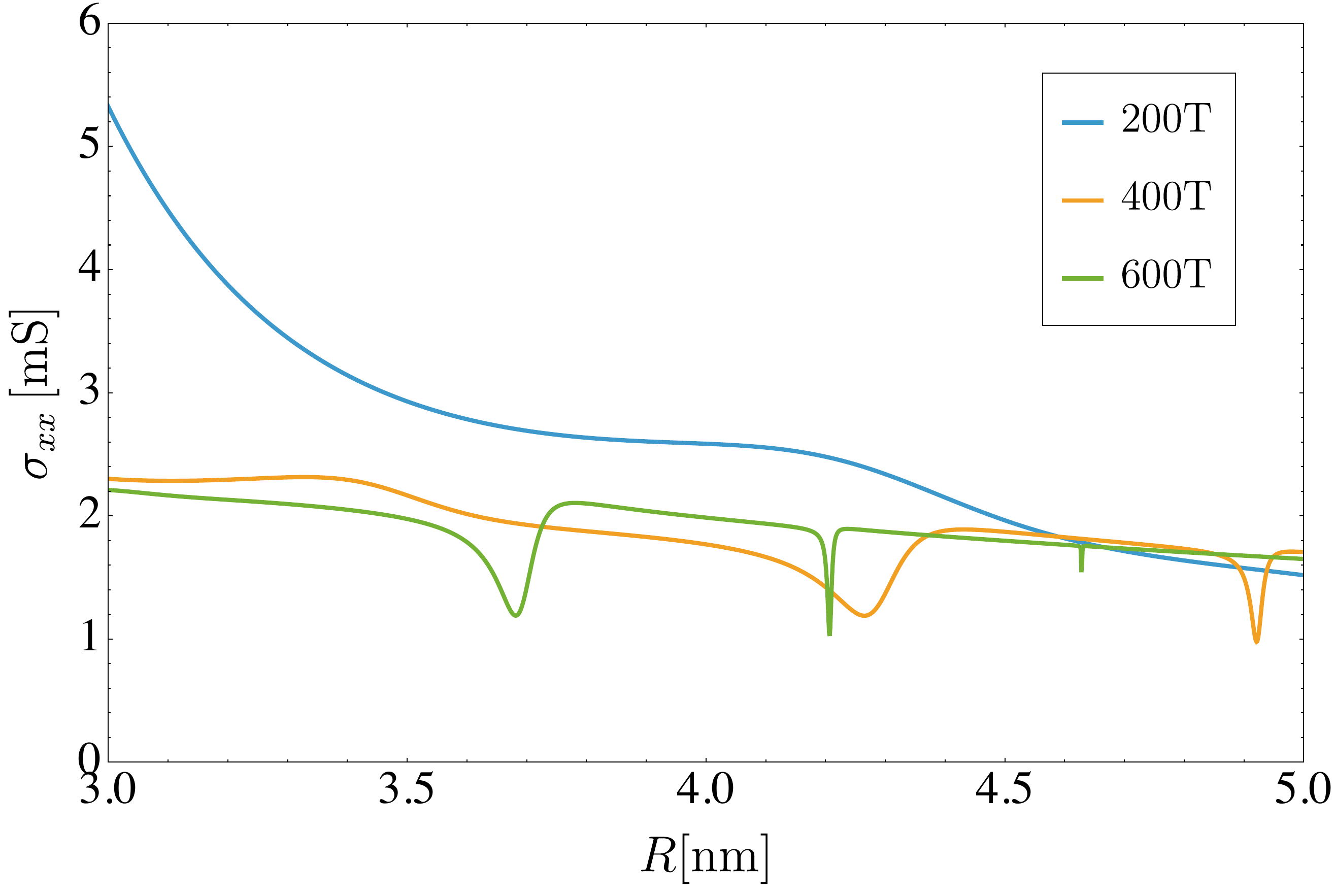}
         \subcaption{}\label{fig:cond_vs_R}
     \end{subfigure}
     \begin{subfigure}[]{0.45\textwidth}
         \centering
         \includegraphics[width=\linewidth]{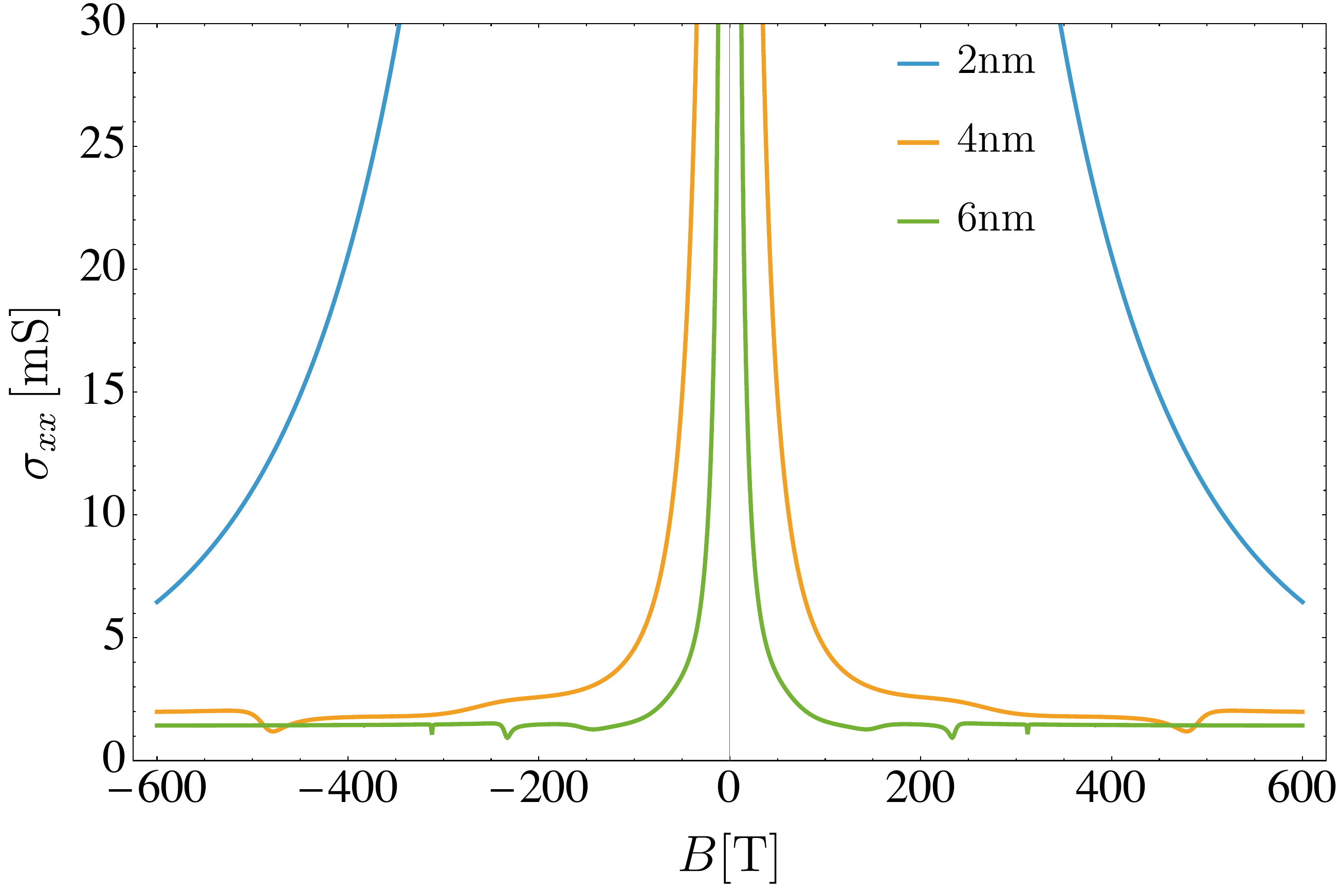}
         \subcaption{}\label{fig:cond_vs_B}
     \end{subfigure}
    \caption{DC conductivity at zero temperature versus (a) the radii of the nanobubbles  and (b) the pseudomagnetic field inside the nanobubbles.} 
    \label{fig:cond_vs_R_B}
\end{figure}

\paragraph{Thermoelectric analysis.} We now turn to the analysis of the electrical conductivity, thermal conductivity, and thermoelectric efficiency. As indicated by Eq.~\eqref{eq:tau_final_2}, the transport relaxation time scales inversely with the nanobubble density in the sample. Consequently, both electrical and thermal conductivities can be tuned by controlling the concentration of nanobubbles. An increase in the nanobubble density by an order of magnitude leads to a corresponding tenfold reduction in both electrical and thermal conductivities.

As shown in Eqs.~\eqref{eq:cond_DC_final} and \eqref{eq:cond_term_final}, the electrical and thermal conductivity tensors are diagonal and isotropic. From this point on, we focus on the $\sigma_{xx}$ and $\kappa_{xx}$ components. The behavior of the conductivity as a function of the radius is shown in Fig.~\ref{fig:cond_vs_R}. The general trend is a slight decrease as the radius increases. Once again, minima appear, which are associated with the discrete spectrum of quasi-bound states within the nanobubbles. 

Figure \ref{fig:cond_vs_B} shows the conductivity as a function of the pseudomagnetic strain field within the nanobubbles. As mentioned earlier, this field couples with opposite signs in each valley, and here we consider the field that couples to the $\chi = +1$ valley. As discussed in the analysis of the role of $\chi$, the conductivity is expected to be symmetric with respect to the sign of the pseudogauge field, as well as for a real magnetic field. The results show that increasing the magnitude of the pseudofield leads to a reduction in conductivity. However, this decrease becomes less pronounced as the strength of the strain field continues to increase.

\begin{figure}[!ht]
    \centering
    \begin{subfigure}[]{0.45\textwidth}
         \centering
         \includegraphics[width=\linewidth]{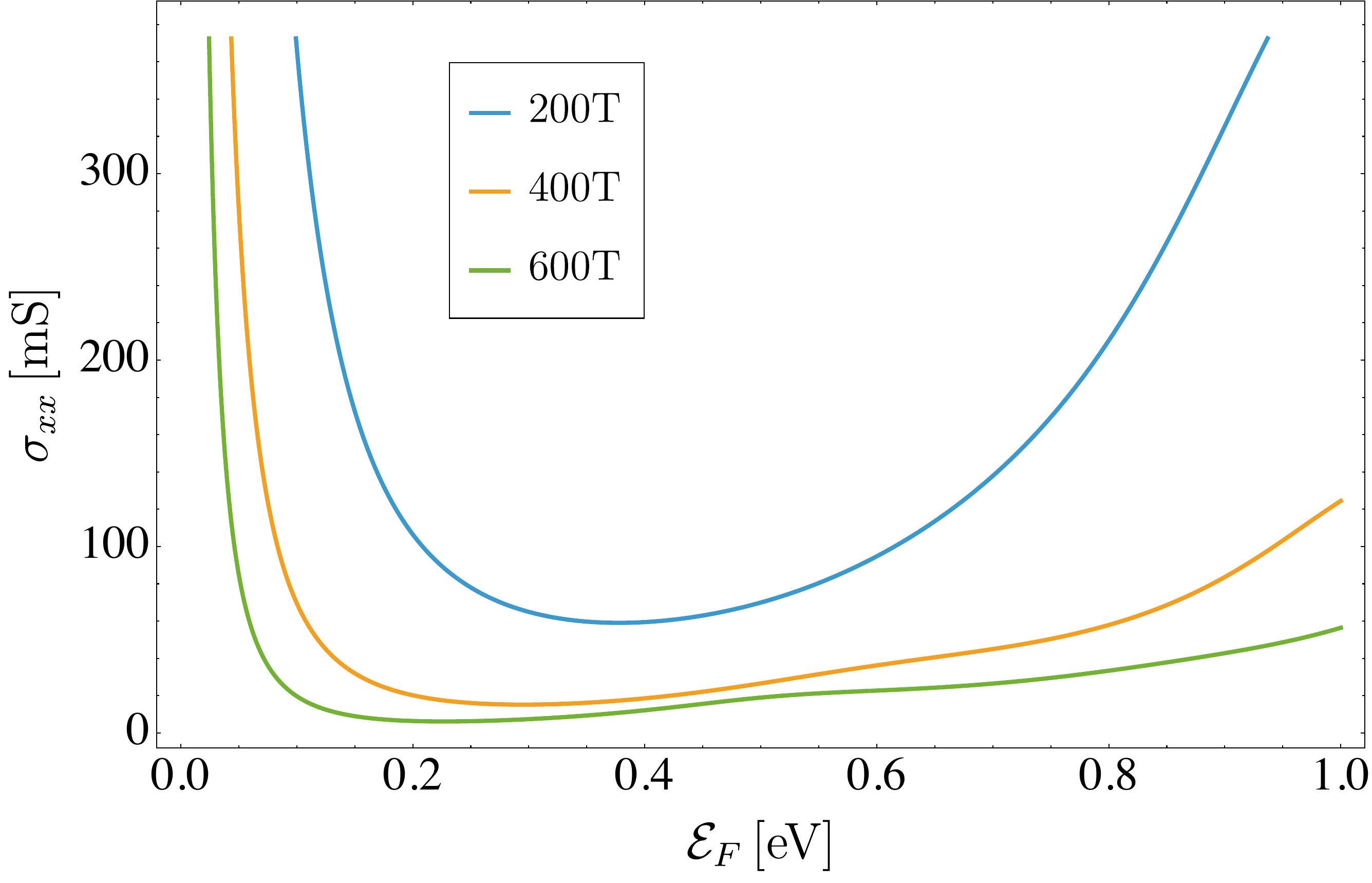}
         \subcaption{}\label{fig:cond_vs_EF_R2}
     \end{subfigure}
     \begin{subfigure}[]{0.45\textwidth}
         \centering
         \includegraphics[width=\linewidth]{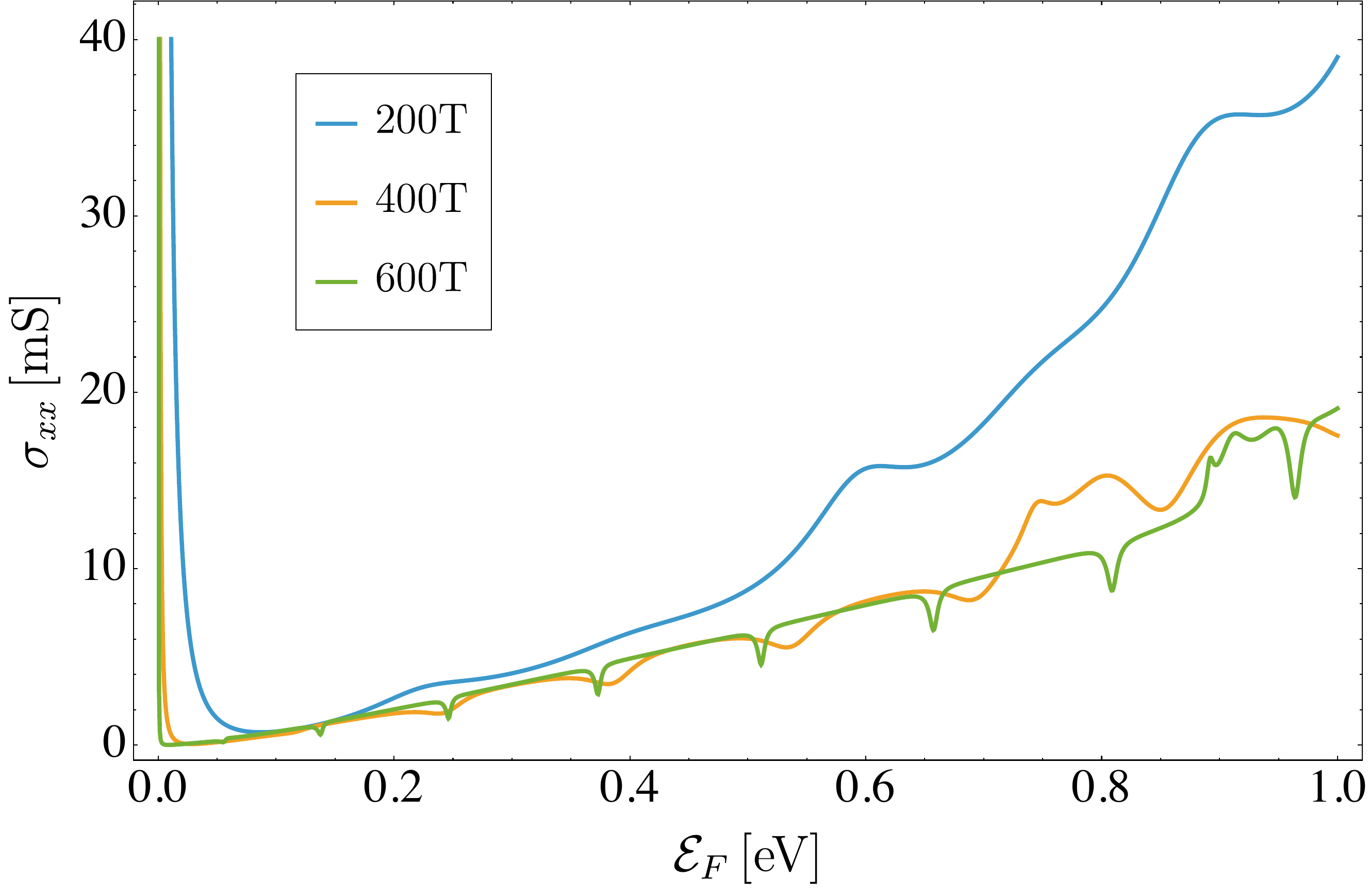}
         \subcaption{}\label{fig:cond_vs_EF_R4}
     \end{subfigure}
      \begin{subfigure}[]{0.45\textwidth}
         \centering
         \includegraphics[width=\linewidth]{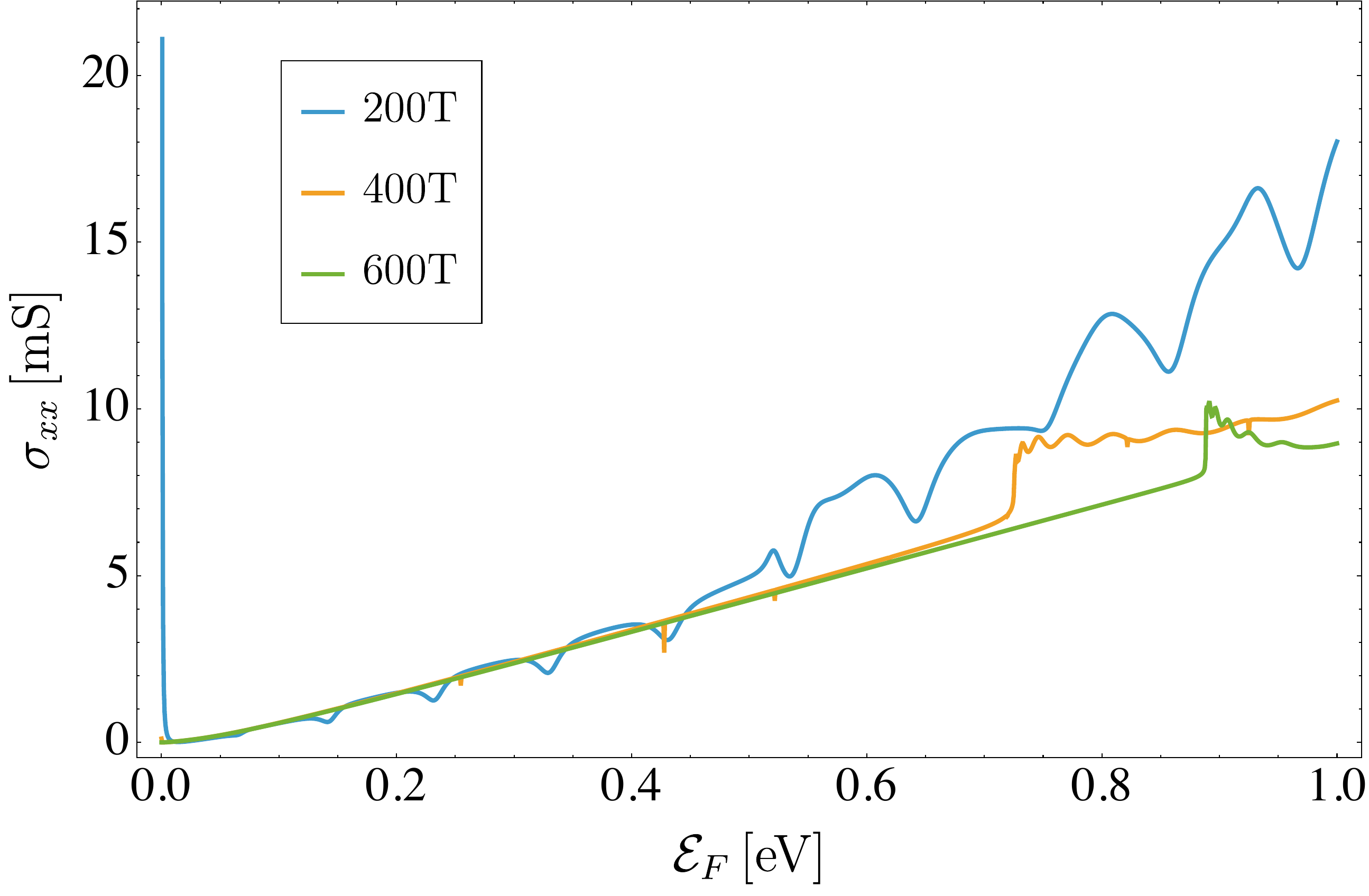}
         \subcaption{}\label{fig:cond_vs_EF_R6}
     \end{subfigure}
    \caption{DC conductivity versus the Fermi energy at zero temperature for a distribution of: (a) 2 nm-radius nanobubbles, (b) 4 nm-radius nanobubbles and (c) 6 nm-radius nanobubbles.} 
    \label{fig:cond_vs_EF}
\end{figure}

Figure \ref{fig:cond_vs_EF} shows the zero-temperature conductivity as a function of the Fermi energy, computed using Eq.~\eqref{eq:cond_DC_zero}, for three representative values of the nanobubble radius. The first notable feature is a divergent trend in the conductivity at low energies for smaller bubble radii and weaker pseudomagnetic fields. This behavior arises because, for small radii and low fields, the number of quasi-bound states that can form is small or even zero, and the region where the strain-induced pseudofield is active becomes so narrow that the scattering centers are effectively transparent to charge carriers. This phenomenon is analogous to the Ramsauer-Townsend effect in the low-energy electron scattering of noble gas atoms \cite{sakurai1994modern}. As can be seen in Figs.~\ref{fig:cond_vs_EF_R2}-\ref{fig:cond_vs_EF_R6}, this effect weakens and tends to disappear as the radius and magnitude of the torsional pseudofield increase. In such cases, scattering becomes significant even at low energies, and the conductivity exhibits the expected behavior: it tends to vanish as the energy approaches zero. As mentioned in the analysis of the relaxation time, the sharp minima in Figs.~\ref{fig:cond_vs_EF_R4}-\ref{fig:cond_vs_EF_R6} indicate the presence of quasi-bound states within the bubbles.

\begin{figure}[!ht]
    \centering
     \begin{subfigure}[]{0.45\textwidth}
         \centering
         \includegraphics[width=\linewidth]{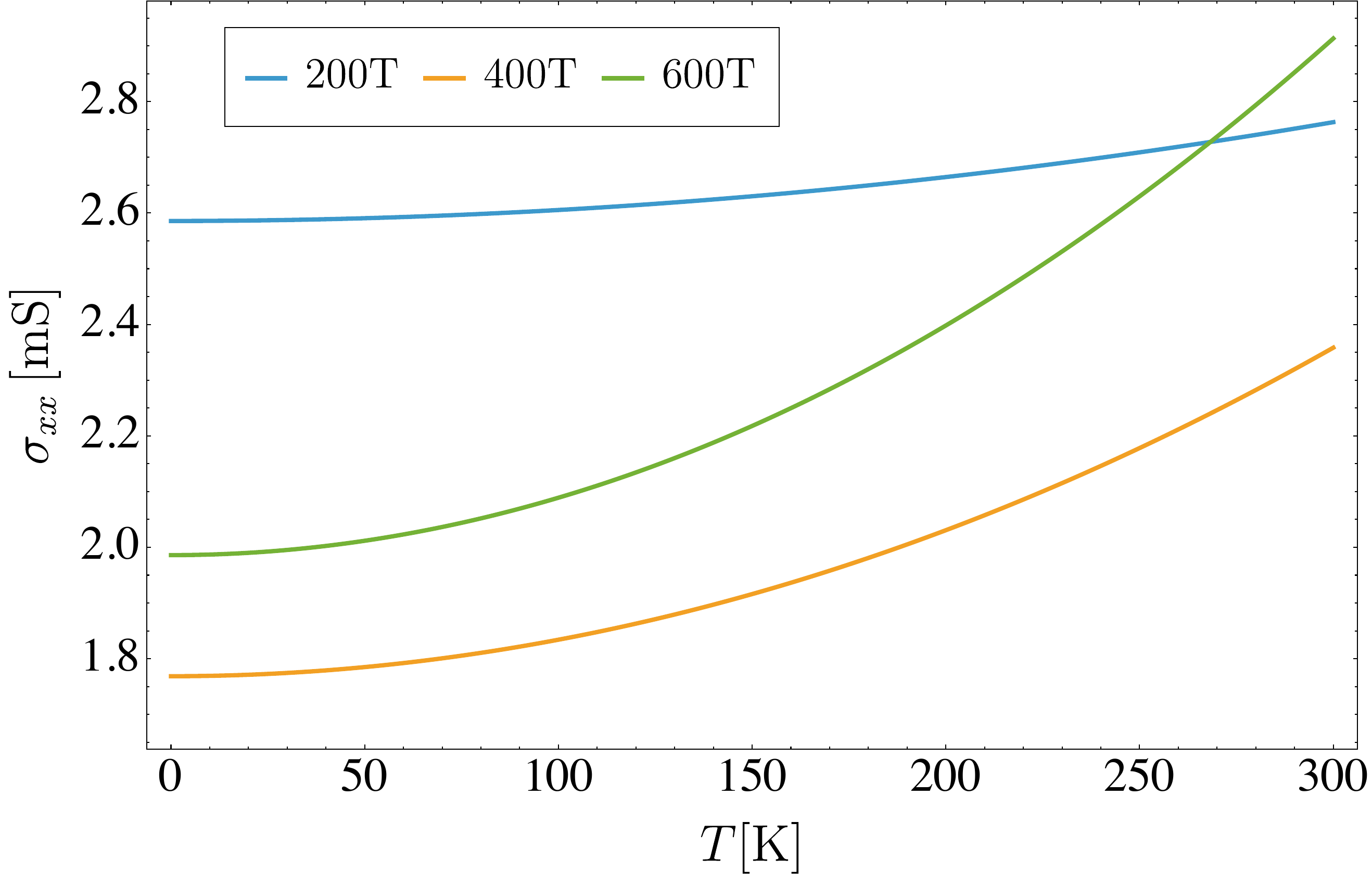}
         \subcaption{}\label{fig:cond_vs_T}
     \end{subfigure}
     \begin{subfigure}[]{0.45\textwidth}
         \centering
         \includegraphics[width=\linewidth]{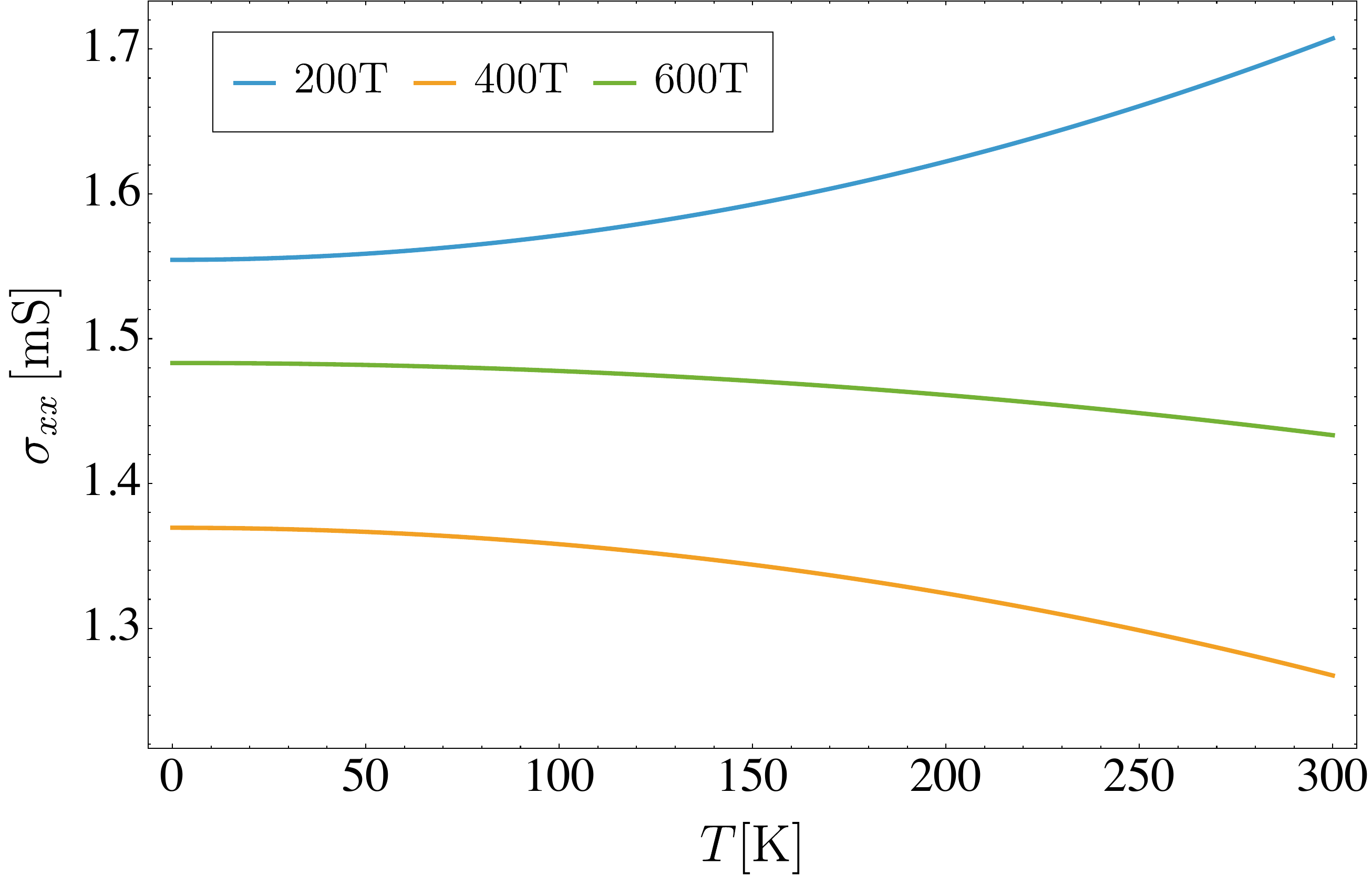}
         \subcaption{}\label{fig:cond_vs_T_EF2}
     \end{subfigure}
    \caption{Temperature dependence of the DC conductivity. The conductivity was calculated for a  density of $n_b = 8 \times 10^{10}$ cm$^{-2}$ of a 4 nm-radius nanobubble distribution, and a Fermi energy of: (a) $\mathscr{E}_F = 200$ meV and (b) $\mathscr{E}_F = 160$ meV.} 
    \label{fig:cond_vs_Temp}
\end{figure}

\subsection*{The role of temperature}

We now analyze how conductivity varies with temperature, as computed from Eq.~\eqref{eq:cond_DC_final} and shown in Fig.~\ref{fig:cond_vs_Temp}. We consider two cases: the first one in which the distribution of bubbles is independent of temperature. This case approaches the real situation in which nanobubbles are formed when graphene is deposited on a substrate that directly induces bubbles with the desired size and distribution, similar to what is reported in Ref.~\cite{S_Levy_2010} for graphene on a platinum substrate or in Ref.~\cite{Lim2013} for the case of graphene on diamond. The second case occurs when the bubbles are generated spontaneously due to thermal effects, and their size and distribution depend on the temperature, as shown in Ref.~\cite{PhysRevB.106.045418}.

\subsubsection{Distribution of bubbles independent of temperature}

 A key point is that, according to the structure of Eq.~\eqref{eq:cond_DC_final}, the conductivity depends on the energy derivative of the relaxation time. As a result, its behavior can be increasing, decreasing, or nearly constant, as illustrated in Figs.~\ref{fig:cond_vs_T}–\ref{fig:cond_vs_T_EF2}. This response is highly sensitive to the value of the Fermi energy at which the conductivity is evaluated. As discussed in the analysis of the relaxation time, this quantity does not always increase monotonically with energy. Figure \ref{fig:tau_vs_EF} clearly shows regions where the relaxation time decreases or remains nearly constant as the energy increases. Therefore, depending on the Fermi energy, the temperature dependence of the conductivity can vary significantly, as illustrated in Fig.~\ref{fig:cond_vs_T_EF2}, which was calculated for a Fermi energy of 160 meV. This implies that the modulation of the electrical conductivity depends strongly on the selected Fermi level, and hence on the carrier density. As a result, it must be analyzed separately for each combination of pseudomagnetic field strength and nanobubble radius.
    \begin{figure}[!ht]
    \centering
    \begin{subfigure}[]{0.45\textwidth}
         \centering
         \includegraphics[width=\linewidth]{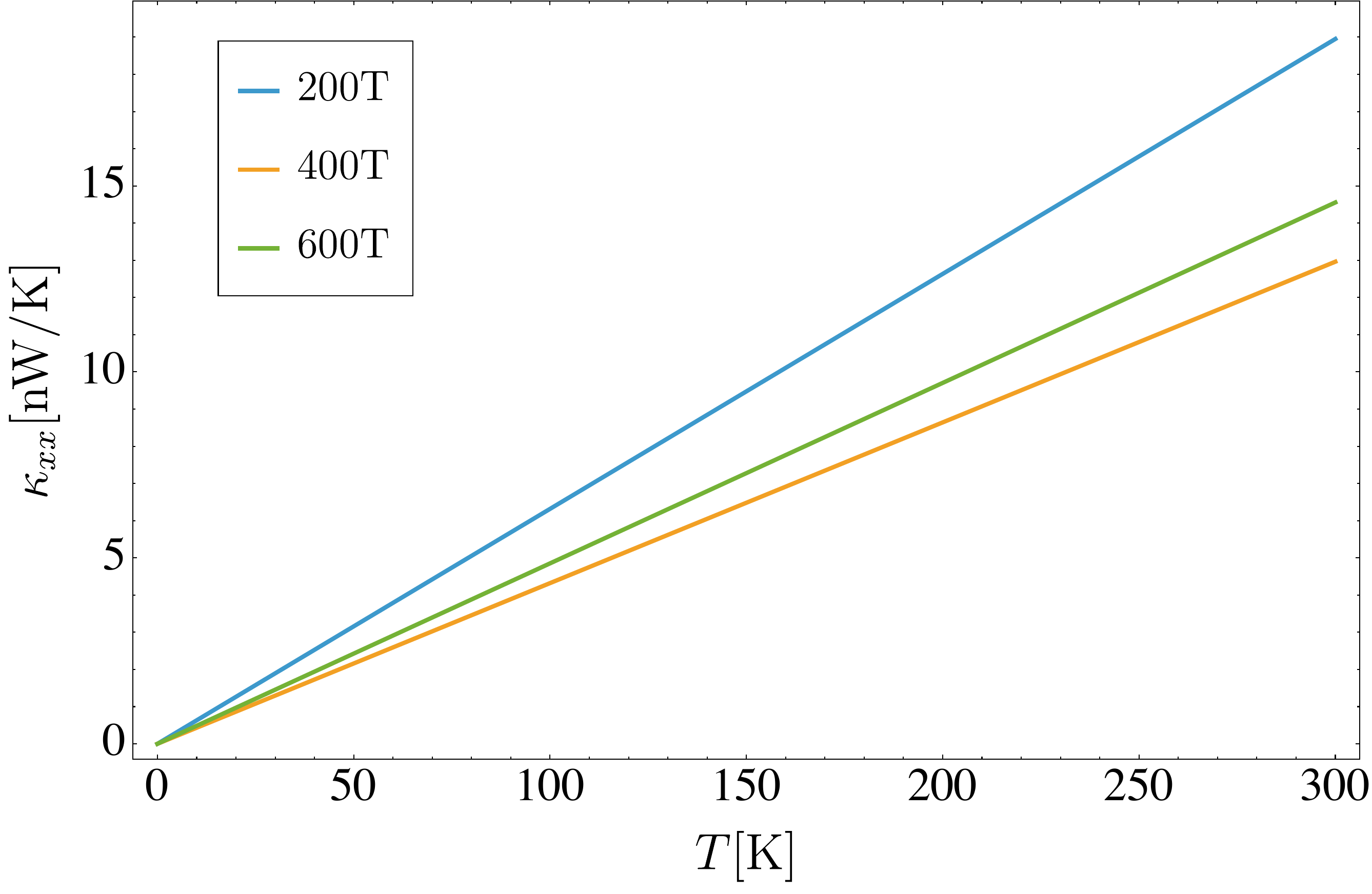}
         \subcaption{}\label{fig:cond_term_vs_T}
     \end{subfigure}
      \begin{subfigure}[]{0.45\textwidth}
         \centering
         \includegraphics[width=\linewidth]{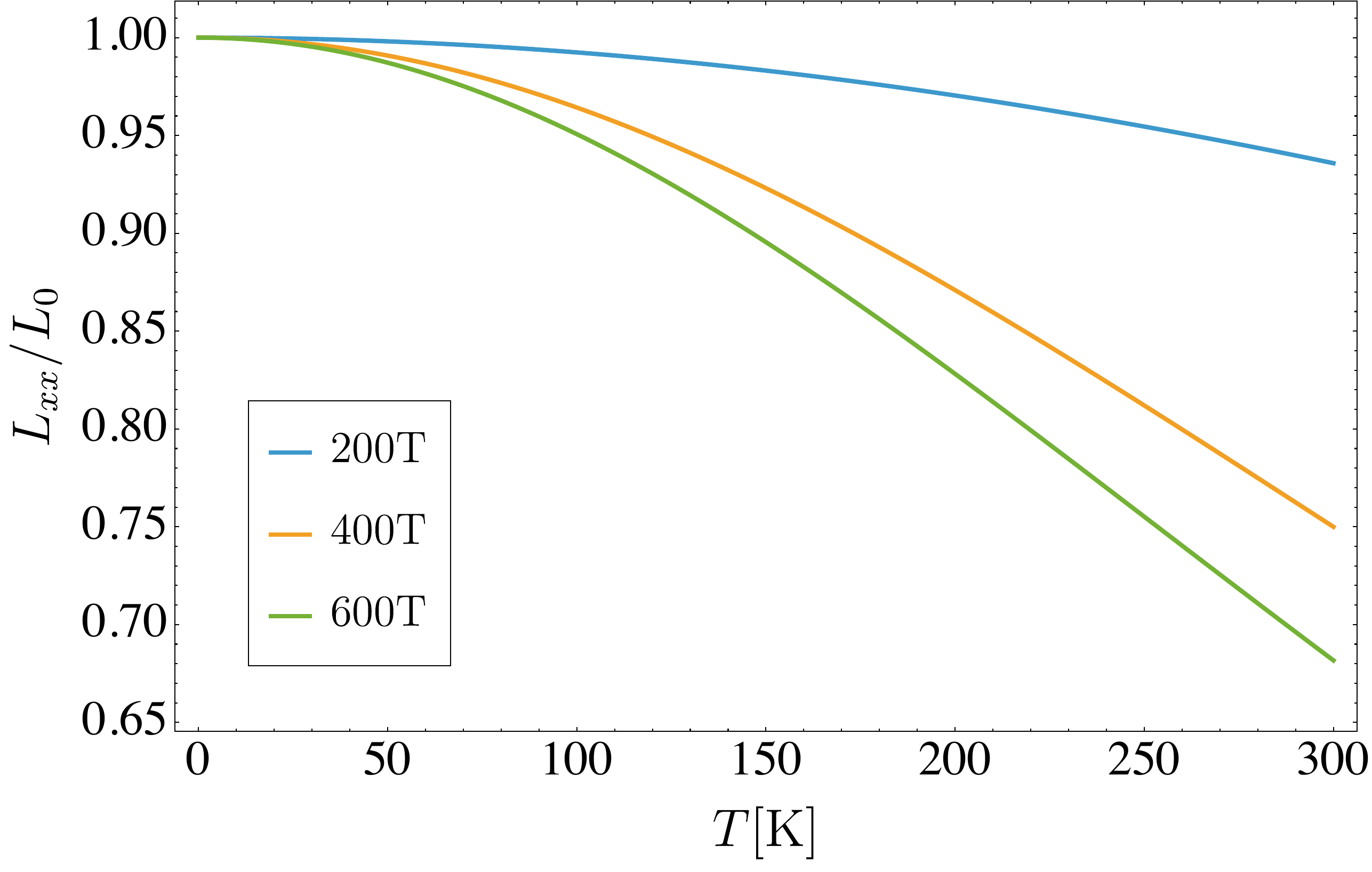}
         \subcaption{}\label{fig:lorenz_vs_T}
     \end{subfigure}
     \caption{Temperature dependence of: (a) the thermal conductivity, and (b) the Lorenz number. The graphs were computed for a nanobubble density of $n_b = 8 \times 10^{10}$ cm$^{-2}$, a nanobubble radius of 4 nm, and a Fermi energy of 200 meV.} 
    \label{fig:thermal}
     \end{figure}
     \begin{figure}[!ht]
    \centering
     \begin{subfigure}[]{0.45\textwidth}
         \centering
         \includegraphics[width=\linewidth]{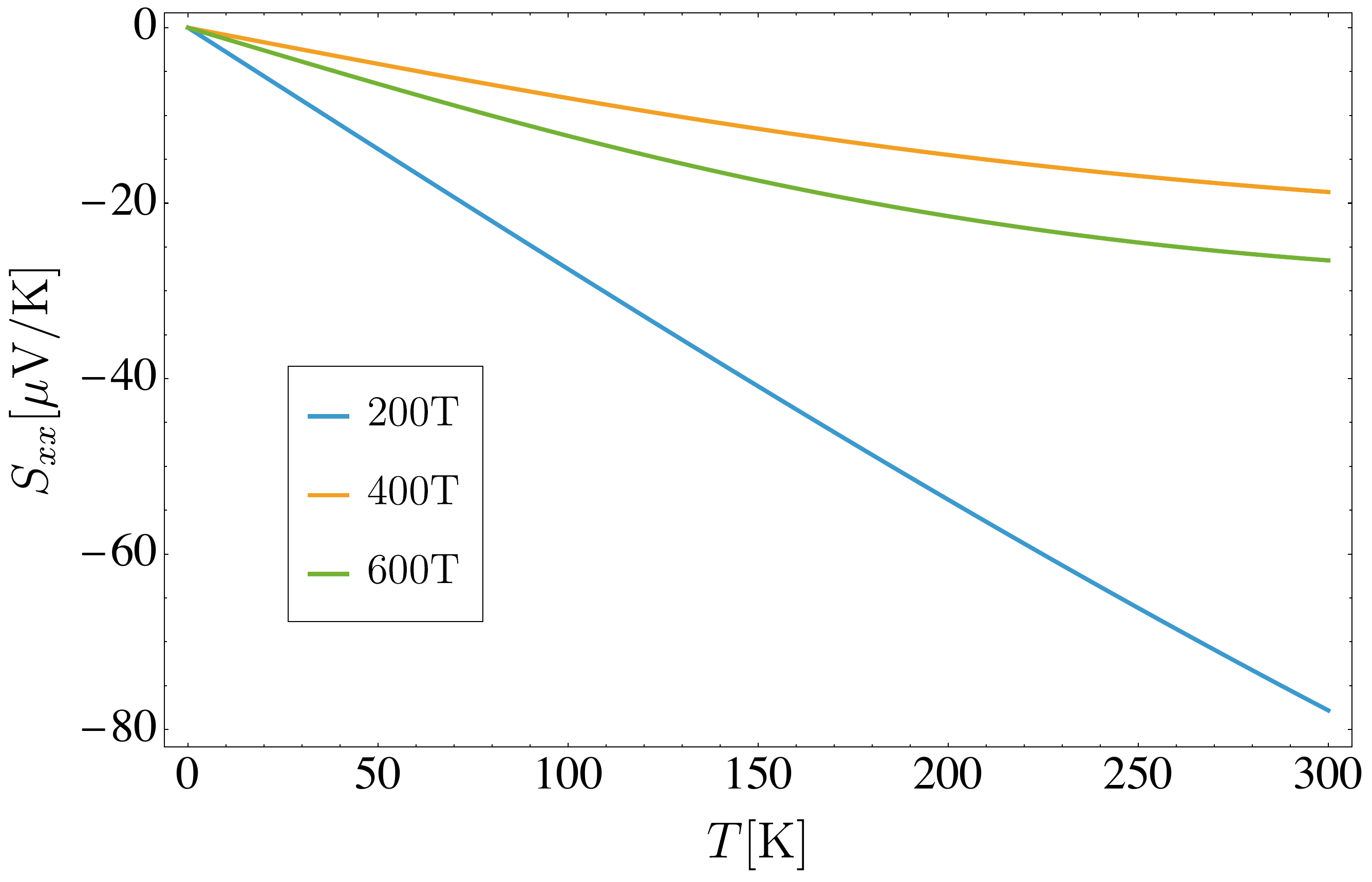}
         \subcaption{}\label{fig:Seebeck_vs_T}
     \end{subfigure}
     \begin{subfigure}[]{0.45\textwidth}
         \centering
         \includegraphics[width=\linewidth]{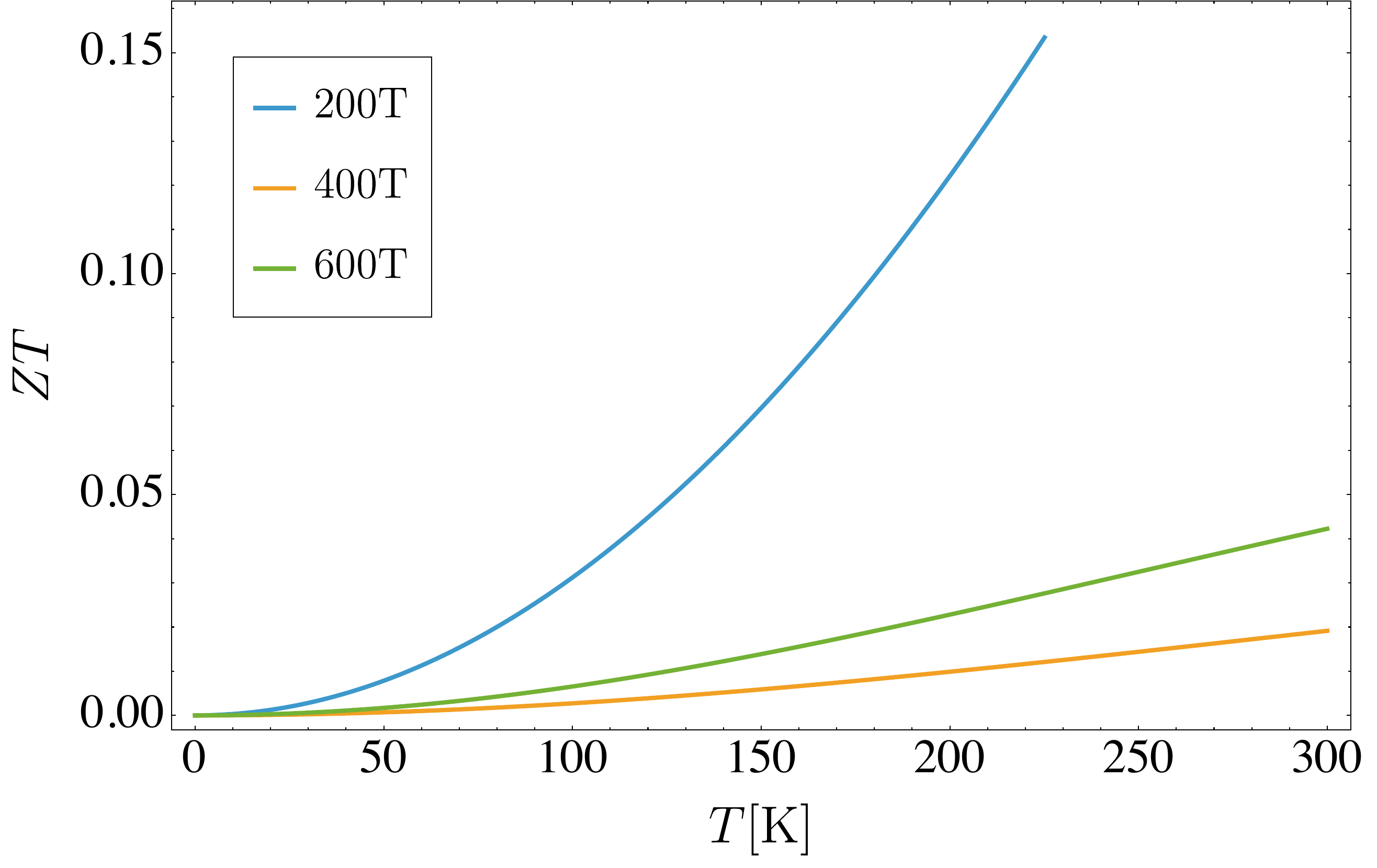}
         \subcaption{}\label{fig:ZT_vs_T}
     \end{subfigure}
    \caption{Temperature dependence of: (a) the Seebeck coefficient, and (b) the figure of merit $ZT$.  The graphs were computed for a nanobubble density of $n_b = 8 \times 10^{10}$ cm$^{-2}$, a nanobubble radius of 4 nm, and a Fermi energy of 200 meV.} 
    \label{fig:Seebeck_ZT}
\end{figure}

Our calculated conductivity values range over approximately two orders of magnitude, from $\sim 1$ mS to $\sim 10^2$ mS, higher for smaller nanobubble radii and lower for larger ones. As mentioned above, the conductivity depends on the nanobubble density and can be tuned by increasing or decreasing their concentration. In the literature, in Ref. \cite{Bahamon} the authors presents a study combining molecular dynamics and tight-binding simulations to analyze how pressure-induced nanobubbles in graphene generate local pseudomagnetic fields of the order of hundreds to thousands of Tesla, capable of confining electrons in graphene nanoribbons. It is shown that these fields induce scattering, mode mixing, and Fano resonances in the conductance, with anti-resonances appearing at the edges of quantum plateaus and sharp resonant peaks at low energies (for electron energies around 0.05 times the hopping parameter $t$). As an example, for an energy of 0.215 $t$, they found the conductivity of an ideal zigzag nanoribbon is approximately eleven times the quantum of conductance ($2e^2/h$), corresponding to 0.852 mS. In the presence of a circular or triangular nanobubble, the conductivity is reduced to approximately 10.1 and 9.4 times $2e^2/h$, i.e., 0.783 and 0.728 mS, respectively. The order of magnitude of these values of the conductivity can be achieved by a concentration of $\sim 10^{12}$ cm$^{-2}$ of a distribution of 4 nm to 6 nm radius nanobubbles.

The thermal conductivity calculated using Eq.~\eqref{eq:cond_term_final} is shown in Fig.~\ref{fig:cond_term_vs_T} as a function of temperature. The obtained values reach up to 20 nW/K, corresponding to approximately $2 \times 10^{-12}$ W/mK, which is fifteen orders of magnitude lower than the values reported in the literature for monolayer graphene, ranging from $\sim$5000 W/mK in experimental measurements ~\cite{Balandin} to 1300 W/mK in first-principles calculations \cite{Han}. It is important to clarify, however, that both the experimental and first-principles results account for many additional scattering mechanisms, particularly electron-phonon and phonon-phonon scattering, which are not included in the present analysis. One must always keep in mind that the total thermal conductivity is the sum of the electronic and lattice contributions, i.e., $\kappa_{t} = \kappa_{e} + \kappa_{l}$. The present results account solely for the contribution due to nanobubbles to the electronic thermal conductivity. Other scattering mechanisms can be incorporated using the well-known Matthiessen’s rule \cite{ziman}. Nonetheless, it has been proposed that the introduction of mechanical defects, such as those induced by strain, can significantly reduce the thermal conductivity of monolayer graphene. For instance, first-principles calculations for kirigami-patterned graphene yield thermal conductivities on the order of $\sim$1–10 W/mK \cite{Gao}.

Figure \ref{fig:lorenz_vs_T} shows the dimensionless ratio $L/L_0$ as a function of temperature, where the Lorenz number $L = \kappa_{xx} / (T \sigma_{xx})$ quantifies the proportionality between the electrical and thermal conductivities. We see that the Wiedemann-Franz law holds at low temperature, because the thermal conductivity in Eq.~\eqref{eq:cond_term_final} is proportional to the residual DC conductivity in Eq.~\eqref{eq:cond_DC_zero}, i.e., $\kappa_{xx}(T)=L_0 T\sigma_{xx}(T= 0)$, where $L_0=\frac{\pi^2}{3}\left(\frac{k_B}{e}\right)^2$. As shown in Fig.~\ref{fig:lorenz_vs_T}, increasing the temperature leads to deviations from metallic behavior, reflected in the departure of the ratio $L/L_0$ from unity. It can also be observed that the deviation from metallic behavior increases with the magnitude of the strain-induced pseudofield.

The Seebeck coefficient, calculated using Eq.~\eqref{eq:Seebeck}, is shown in Fig.~\ref{fig:Seebeck_vs_T} as a function of temperature. It quantifies the voltage difference generated by a temperature gradient, indicating how efficiently a material converts heat into electricity. The negative sign reflects that the charge carriers are electrons, which have negative charge. The general trend is an increase in magnitude with temperature and a decrease in magnitude with increasing pseudomagnetic field. The obtained values, reaching up to 80 $\mu$V/K, are of the same order of magnitude as those reported in the literature, typically around 40–60 $\mu$V/K~\cite{HWANG2023467}.

Finally, the dimensionless figure of merit $ZT = S_{xx}^2 T \sigma_{xx} / \kappa_{xx}$ is shown in Fig.~\ref{fig:ZT_vs_T} as a function of temperature. The figure of merit quantifies the efficiency with which a material converts heat into electricity, accounting for how well it conducts electricity, how poorly it conducts heat, and how large a voltage it generates in response to a temperature gradient. A $ZT$ value close to 1 is considered good for practical applications, while values above 2 are regarded as excellent, enabling highly efficient thermoelectric devices. Our results can reach values above $ZT\sim0.15$, with a general trend of increasing with temperature and decreasing with the magnitude of the strain-induced pseudofield. Reported values of the electronic figure of merit $ZT$ for graphene nanoribbons range from 0.01 to 0.5~\cite{ZT_1}, and it has been shown that, by combining chevron-type geometry and isotope engineering, the $ZT$ value can be enhanced up to 2.45 at 800 K~\cite{ZT_2}. Our results do not show a significant improvement in thermoelectric efficiency; however, larger nanobubbles may lead to greater enhancement. The nanobubble concentration does not affect the behavior of the Seebeck coefficient or the figure of merit, since both quantities are computed from ratios involving the transport time or its energy derivatives, which scale inversely with the bubble density, leading to a cancellation of the effect. \\

\subsubsection{Temperature dependent distribution of bubbles}

As mentioned in Ref.~\cite{PhysRevB.106.045418}, in graphene, thermal fluctuations at finite temperature $T$ drive the formation of out-of-plane ripples. In the circular model, a ripple can be considered as a nanobubble of height $t$ and an effective diameter $D$. At room temperature most circular ripples lie around $D\sim5$–$12$\AA. As $T$ increases the peaks in the $D$-distribution broaden: for $D<8\,$\AA\ the density rises with $T$, for $D>9\,$\AA\ it initially falls, and near $D\approx8.7\,$\AA\ it is nearly $T$-independent. Overall, higher $T$ ultimately promotes the appearance of larger ripples that dominate above room temperature. According to Ref.~\cite{PhysRevB.106.045418}, the circular-model ripple distribution depends on the ratio $t/D$ as
\begin{align}
n_b(t,D) = A(T) \exp \left[-C(T)\frac{\lambda}{k_B T}\left(\frac{t}{D}\right)^2\left(\frac{D}{L_{\mathrm{th}}}\right)^{\eta}\right],\label{eq:distribution_bubbles}
\end{align}
where $A(T)$ is a temperature dependent normalization factor, $C(T)$ is a dimensionless fitting parameter, $\lambda$ is the bare bending rigidity fixed by $\lambda/k_B T\approx 40$ for $T=300$ K, $k_B$ is Boltzmann's constant, and $\eta\simeq 0.8$ is the critical exponent. $L_{\mathrm{th}}$ is a critical length, which in graphene is $\approx 3$ nm, and above which Eq.~\eqref{eq:distribution_bubbles} is valid. 

\begin{figure}[!ht]
    \centering    \includegraphics[width=0.9\linewidth]{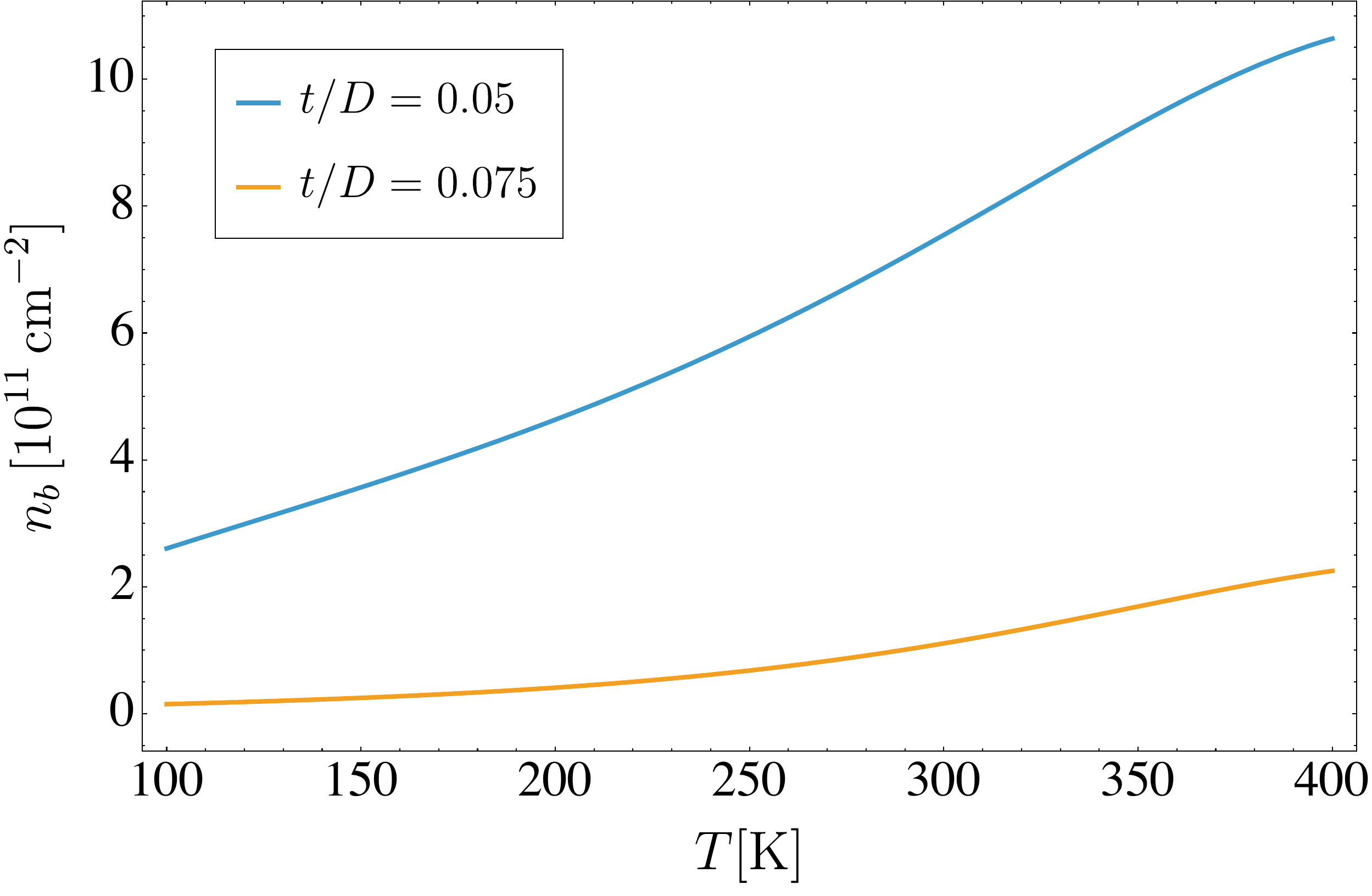}
    \caption{Temperature dependence of the nanobubble density in graphene, obtained from Eq.~\eqref{eq:distribution_bubbles} using the fitted parameters $A$ and $C$ extracted from the data reported in Ref.~\cite{PhysRevB.106.045418}. We changed the units from $10^3\,\mu\text{m}^{-2}$ to $10^{11}$ cm$^{-2}$.}
    \label{fig:n_vs_T}
\end{figure}

Reference \cite{PhysRevB.106.045418} provides a table with values for the parameters $A$ and $C$ at temperatures of 100, 200, 300, and 400 K. $A(T)$ is given in units such that the ripple density is expressed in $10^{3}\,\mu\text{m}^{-2}$. Based on these values, we numerically extrapolated the behavior of $A$ and $C$ as a function of temperature. We selected two $t/D$ ratios: 0.05 and 0.075. This choice was made because, as clearly shown in Ref.~\cite{PhysRevB.106.045418}, the ripple density decreases drastically when this ratio increases. The behavior of the nanobubbles (circular ripples) density as a function of temperature is presented in Fig.~\ref{fig:n_vs_T}, where it is evident that the density decreases by approximately one order of magnitude when going from $t/D=0.05$ to $t/D=0.075$. Moreover, the general trend for these $t/D$ ratios is a monotonic increase with increasing temperature.

\begin{figure}[!ht]
    \centering
     \begin{subfigure}[]{0.45\textwidth}
         \centering
         \includegraphics[width=\linewidth]{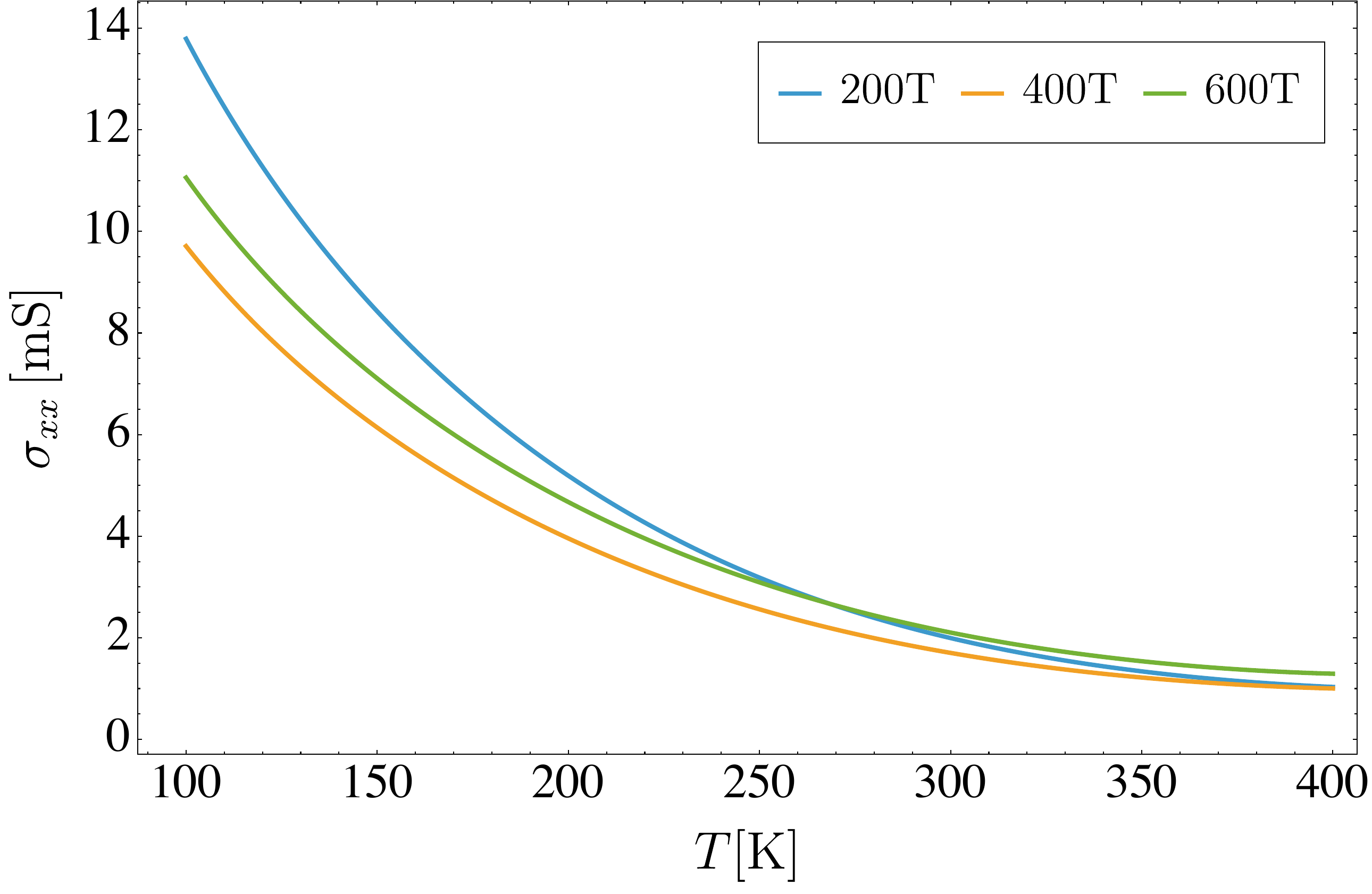}
         \subcaption{}\label{fig:cond_vs_T_2}
     \end{subfigure}
     \begin{subfigure}[]{0.45\textwidth}
         \centering
         \includegraphics[width=\linewidth]{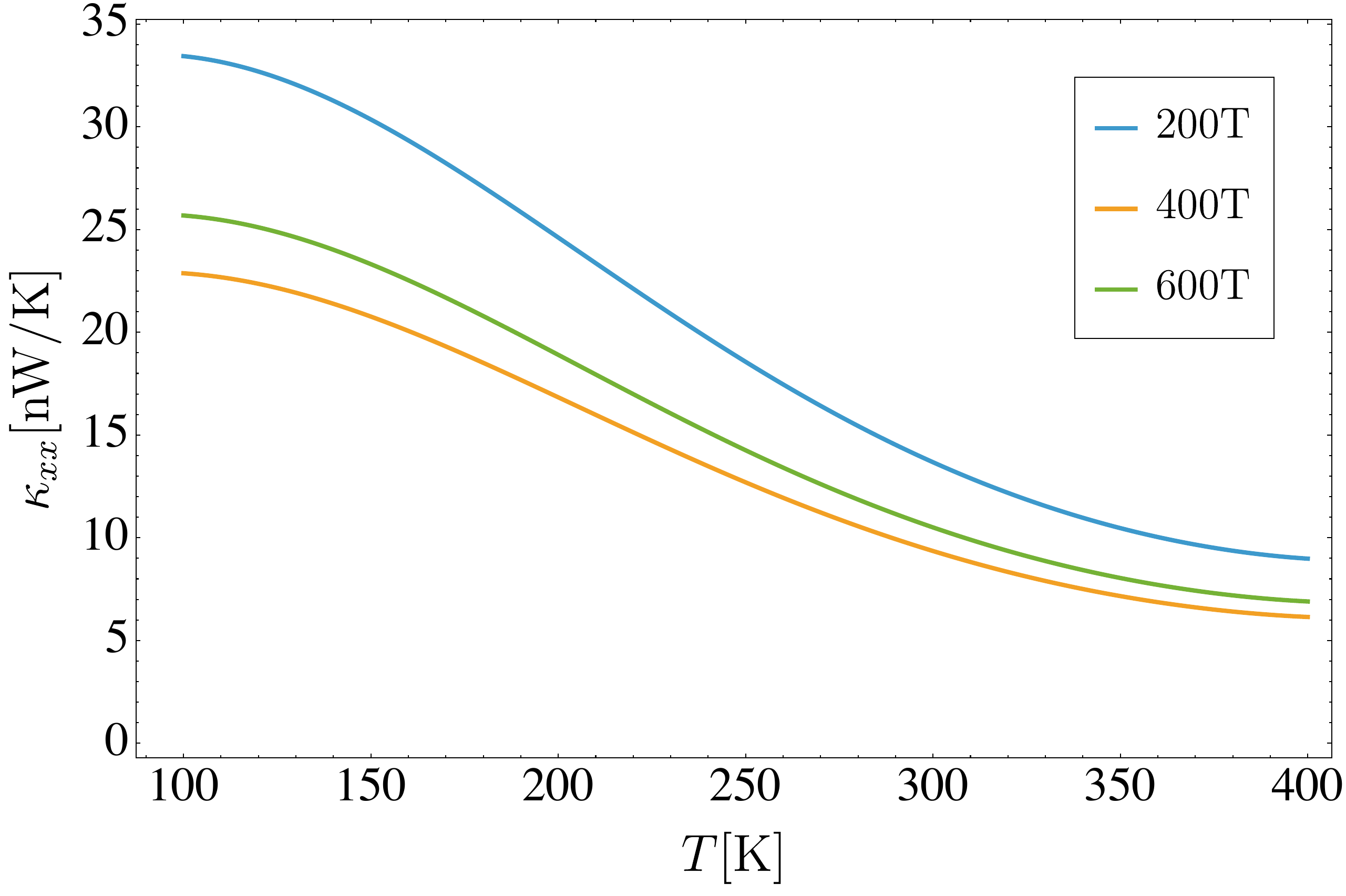}
         \subcaption{}\label{fig:cond_term_vs_T_2}
     \end{subfigure}
    \caption{Temperature dependence of: (a) the DC conductivity, and (b) the thermal conductivity computed taking into account the temperature dependence of the distribution of nanobubbles and for a ratio $t/D=0.075$.} 
    \label{fig:conductivities_2}
\end{figure}

The electrical and thermal conductivities, calculated using Eq.~\eqref{eq:cond_DC_final} and Eq.~\eqref{eq:cond_term_final}, respectively, and considering a nanobubble density dependent on temperature in the relaxation time of Eq.~\eqref{eq:tau_final_2}, are shown in Fig.~\ref{fig:conductivities_2}. It is important to note that the monotonically increasing trend found in the case where the bubble distribution is temperature-independent is reversed. In this case, both the electrical and thermal conductivities decrease with increasing temperature, since, as mentioned earlier, the ripple density increases with temperature.

Regarding the thermoelectric efficiency, the values of the Seebeck coefficient $S_{xx}$, the Lorenz number $L/L_{0}$, and the figure of merit $ZT$ do not change, since in our model we have assumed that the transport relaxation time is inversely proportional to the nanobubble density, as shown in Eq.~\eqref{eq:tau_final_2}. This implies that both the DC and thermal conductivities are also inversely proportional to $n_b(T)$, and since $S_{xx}$, $L/L_{0}$, and $ZT$ are defined from the ratio between $\sigma_{xx}$ and $\kappa_{xx}$, the dependence on $n_b$ is canceled out. This means that, within this model, the thermoelectric performance is determined solely by the scattering mechanism itself and not by the number of scattering centers.

\section{Conclusions}
\label{Conclusions}

In this work we have investigated the thermoelectric transport properties of monolayer graphene in the presence of randomly distributed strain fields, modeled as localized perturbations via a two-dimensional Dirac oscillator formalism.  We began by introducing the theoretical framework that relates strain-induced modifications in the hopping amplitudes to an effective pseudogauge field, which acts analogously to a magnetic vector potential in the low-energy Dirac theory of graphene. For specific strain configurations, this pseudogauge field yields a constant pseudomagnetic field within localized regions, justifying the use of a Dirac oscillator as a minimal and tractable model for the scattering centers. 

Building on this foundation, we developed a semiclassical kinetic theory to compute the thermoelectric transport coefficients, incorporating the full quantum mechanical scattering problem via the $T$-matrix formalism. The transport relaxation time was obtained analytically by solving the scattering problem in the partial wave expansion, using the exact solutions of the Dirac oscillator inside the nanobubble region and matching them to plane-wave solutions outside. The phase shifts, which exhibit a strong dependence on both the Fermi energy and the pseudomagnetic field, encode the resonant structure of the problem and control the transport time through quantum interference effects. Importantly, the valley symmetry of the relaxation time was preserved due to a nontrivial symmetry in the phase shifts under inversion of the valley index and angular momentum.

In the subsequent analysis, we derived closed-form expressions for the DC electrical conductivity, Seebeck coefficient, and electronic thermal conductivity, incorporating the Sommerfeld expansion to capture finite temperature effects. Our numerical evaluation revealed that the conductivity and thermal conductivity can be effectively tuned by varying the nanobubble radius, density, and the strength of the pseudomagnetic field. Notably, we observed pronounced dips in the relaxation time and conductivity spectra, attributable to quasi-bound states associated with the confined pseudo-Landau levels, a signature of resonant scattering. These features cannot be captured by the Born approximation and underscore the necessity of the full $T$-matrix treatment using phase shifts.

At low energies and weak pseudofields, the conductivity displayed a divergent behavior reminiscent of the Ramsauer-Townsend effect, suggesting near-transparent scattering centers under certain conditions. As the strain field increases, these transparent regimes vanish, and the system enters a regime dominated by resonant scattering. The temperature dependence of the transport coefficients showed a nontrivial interplay with the energy dependence of the relaxation time, with regimes where the conductivity increases, decreases, or remains flat with temperature depending sensitively on the Fermi level.

Thermoelectric efficiency was assessed through the Lorenz number, the Seebeck coefficient, and the figure of merit $ZT$. While the Wiedemann-Franz law held at low temperatures, deviations emerged at higher temperatures, signaling a breakdown of metallic behavior. The Seebeck coefficient reached magnitudes consistent with previously reported experimental values, and the calculated electronic figure of merit $ZT$ reached values up to 0.15, within the range of reported theoretical predictions for graphene nanoribbons. Although not sufficient for high-efficiency thermoelectric applications on its own, our results suggest that further enhancement is possible through optimized strain profiles and nanobubble radii. 

Finally, by incorporating the temperature dependence of the nanobubble distribution into the relaxation time, we find that the increase in ripple density with temperature leads to a suppression of both electrical and thermal conductivities, in contrast to the monotonically increasing trends obtained under a temperature-independent distribution. Nevertheless, the thermoelectric coefficients $S_{xx}$, $L/L_{0}$, and $ZT$ remain unaffected, since their definitions involve the ratio between $\sigma_{xx}$ and $\kappa_{xx}$, which removes the explicit dependence on $n_b(T)$. These results indicate that, within this framework, thermoelectric performance is governed primarily by the nature of the scattering mechanism, rather than by the absolute number of scattering centers.

\section*{Acknowledgments}
J.A.C. was supported by the SECIHTI PhD fellowship No. 933498. D.A.B. was supported by the DGAPA-UNAM Posdoctoral Program.  A.M.-R. acknowledges financial support by DGAPA-UNAM project No. IG100224, by SECIHTI project No. CBF-2025-I-1862 and by the Marcos Moshinsky Foundation. The authors thank the referees for the insightful comments and suggestions which help to improve the quality of the paper.

%------------------------------------------------------

\appendix

\bibliography{Refs.bib}
\bibliographystyle{unsrt}
\end{document}